%% file: main_Arxiv_v2.tex
\documentclass[journal=apchd5,manuscript=draft]{achemso}

\usepackage[T1]{fontenc}
\usepackage[english]{babel}
\usepackage{amsmath,amsthm,amssymb,latexsym,amsfonts,gensymb} 
\usepackage[printonlyused,nolist]{acronym} 
\usepackage{graphicx,float,epstopdf,hyperref} 

\usepackage{xcolor}
\usepackage{soul}
\usepackage[normalem]{ulem}

\graphicspath{{../figures}}
\title{Polarization-controlled strong light--matter
interaction with templated molecular aggregates}

\author{Roland~Sch\"afer}
\affiliation{Institute for Light and Matter, Department f\"ur Chemie, Universit\"at zu K\"oln, Greinstr. 4--6, 50939 K\"oln, Germany}
\email{roland.schaefer@uni-koeln.de}

\author{Philipp~Weitkamp}
\affiliation{Institute for Light and Matter, Department f\"ur Chemie, Universit\"at zu K\"oln, Greinstr. 4--6, 50939 K\"oln, Germany}

\author{Otgonbayar~Erdene-Ochir}
\affiliation{Institute for Light and Matter, Department f\"ur Chemie, Universit\"at zu K\"oln, Greinstr. 4--6, 50939 K\"oln, Germany}

\author{Klaus~Meerholz}
\affiliation{Institute for Light and Matter, Department f\"ur Chemie, Universit\"at zu K\"oln, Greinstr. 4--6, 50939 K\"oln, Germany}

\author{Klas~Lindfors}
\affiliation{Institute for Light and Matter, Department f\"ur Chemie, Universit\"at zu K\"oln, Greinstr. 4--6, 50939 K\"oln, Germany}
\email{klas.lindfors@uni-koeln.de}


\begin{document}
	\begin{acronym}
		\acro{AOI}[AOI]{angle of incidence}
		\acrodefplural{AOI}{angles of incidence}
		\acro{PVD}[PVD]{physical vapor deposition}
		\acro{PL}[PL]{photoluminescence}
        \acro{TDM}[TDM]{transition dipole moment}
		\acro{GNR}[7-AGNRs]{seven-atom wide armchair-edge graphene nanoribbons}
	\end{acronym}

\begin{abstract}

We demonstrate strong light-matter interaction for a layer of templated merocyanine molecules in a planar microcavity. Using a single layer of graphene nanoribbons as a templating layer, we obtain an aligned layer of aggregated molecules. The molecular layer displays anisotropic optical properties resembling those of a biaxial crystal. The anisotropic excitonic component in the cavity results in strongly polarization-dependent light-matter interaction and in increased Rabi-energies. The increased light-matter interaction is possibly due to reduced molecular disorder in the templated molecular layer. This conclusion is supported by an analysis based on a multi-oscillator model. We further use photoluminescence microspectroscopy to demonstrate that the light-matter coupling is spatially homogeneous. Our study introduces molecular templating to strong light-matter studies. The reduced disorder of the system as a consequence of templating is highly beneficial for engineering light-matter interaction.

\end{abstract}

\section*{Introduction}

Light-matter interaction of organic materials in planar microcavities has been the topic of intense research in recent years.\cite{Lidzey1998,Zasedatelev2019,Kena-Cohen2010,Lerario2017,Daskalakis2014,Plumhof2014,Mischok2023,Orgiu2015,berghuis2022,Mischok2024} Organic semiconductors offer large oscillator strengths and materials with transition wavelengths from the ultraviolet to the near-infrared. Among organic materials, Frenkel excitons in J-aggregated molecules show intense and narrowband absorption~\cite{Wurthner2011,Fidder1990,Kobayashi1996} making such materials highly interesting for studies of light-matter interaction and even reaching the strong-coupling regime~\cite{Lidzey1998}. In the strong-coupling regime, the energy exchange rate between light and the excitonic component must exceed all other loss mechanisms so that a periodic exchange of energy can take place.~\cite{garcia-vidal:2021} 
Further, the system must be described using hybrid light-matter states, polaritons.~\cite{khazanov2023} Polaritonic systems based on organic materials have been explored in a variety of applications and effects such as transistors~\cite{Zasedatelev2019}, lasing~\cite{Kena-Cohen2010}, polariton condensates~\cite{Lerario2017,Daskalakis2014,Plumhof2014}, light-emitting organic diodes with angle-independent emission~\cite{Mischok2023}, enhanced charge transport~\cite{Orgiu2015} and exciton propagation~\cite{berghuis2022}, and optical filtering~\cite{Mischok2024}. Key to these developments is maximizing the strength of the light-matter interaction.

The organic systems used in strong light-matter coupling studies are typically amorphous and disordered. The disorder can be static, e.g., positional or orientational, or dynamic resulting from interaction between excitations and vibrational modes. Static disorder results in spectral broadening counteracting the formation of spectrally sharp transitions in for example J-aggregated systems. Orientational disorder further makes it impossible to use the polarization of light as a control-parameter in engineering light-matter coupling. Disorder has more subtle consequences for the energy level structure of the coupled system. The excitonic system is commonly represented as an ensemble of $N$ two-level systems. Coupling this to the near-resonant cavity mode results in an upper (UP) and lower (LP) polariton, which are the bright states of the system, and $N-1$ dark excitations. In the presence of disorder, this picture is changed, and the previously dark states become optically allowed.~\cite{adom.202302387,khazanov2023,gera2022} They are thus often termed "gray states". Beyond linewidth broadening or isotropic optical response, disorder has been predicted to affect the dynamics~\cite{engelhardt2022,agranovich2003} and coherence~\cite{litinskaya2006} of the polaritonic system. On the other hand, strong light-matter coupling has also been shown in theoretical studies to enhance the coherence length of disordered J-aggregated systems.~\cite{Spano2015} For controlled studies and applications it would be desirable to control and maximize the order of the organic semiconductor.

One approach for ordered organic semiconductors is the use of single crystalline materials. This has been successfully done in several studies.~\cite{berghuis2022,kena-cohen2008,berghuis2022b} Obtaining single crystals of the molecules of interest is in most cases not trivial, and integrating the crystalline material into device structures is complicated and in many cases not compatible with fabrication processes. Alternative approaches for reducing disorder have therefore been studied. On single crystalline metal surfaces organic semiconductors are known to form highly ordered thin films, whose crystal structure often deviates from the bulk structure.~\cite{Gaertner2014,molecules26082393,Guo_2014,OCAL2025122690} This high interaction-energy driven effect has been observed for the merocyanine dye 2-[5-(5-dibutylamino-thiophen-2-yl-methylene)-4-\textit{tert}-butyl-5\textit{H}-thiazol-2-ylidene]-malononitrile  (from hereon referred to as HB238\cite{buerckstuemmer2011})
on a Ag(100) surface, resulting in the formation of a chiral tetramer driven by a high Ag-S interaction strength.~\cite{D3NR00767G} The periodicity of single-layered two-dimensional materials such as graphene or hexagonal boron nitride has been used to replace single crystalline metal templates. The growth of organic crystalline layers on those materials is often governed by the formation of needle networks following the six- or three-fold symmetry of the underlying template.~\cite{Kratzer_2019,Hlawacek_2013,Kratzer_2016} Reduced symmetry films showing one preferential direction of crystal growth have been reported for solution-based techniques like shear force crystallisations which have been shown to result in optically anisotropic organic thin films.~\cite{Giri_2014,D4TC00678J,shaw_2016,Herrmann2024} Further uniaxial aligned thin films for cavity applications were realised using liquid crystalline polymer films on a sulfuric dye photoalignment layer.~\cite{LeRoux2020} Here we use an aligned layer of \ac{GNR} as a template to align and order an organic semiconductor in a microcavity. Bottom-up grown graphene nanoribbons are an atomically precise material of reduced dimensionality.~\cite{groening2018,rizzo2018,ruffieux2016,llinas2017,ruffieux2012,denk2014,cai2010,senkovskiy2017} The properties of the ribbons and thus the templating can be engineered through the choice of the precursor molecule used for the growth of the material. The presence of the templating layer results in strongly anisotropic optical properties of the organic excitonic component whose influence on light-matter coupling we characterize in detail. The use of GNRs for templating is potentially a broadly applicable approach to influence the structure of organic semiconductors and not restricted to specific molecules.

\section*{Results and discussion}

\subsection*{Templated thin films}

\begin{figure}
    \centering
    \includegraphics[width=15.5cm]{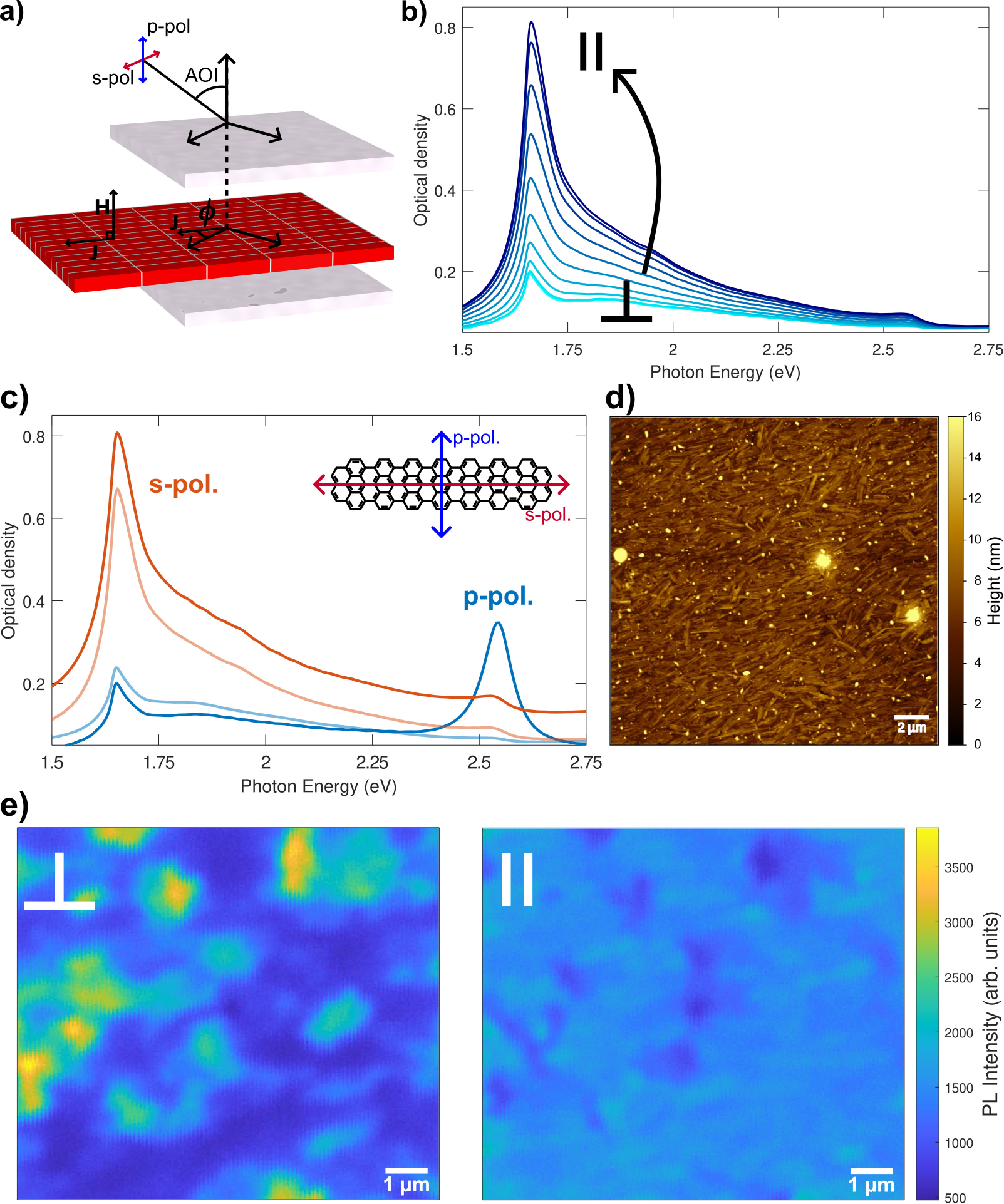}
    \caption{Molecular alignment due to templating with graphene nanoribbons. a) A layer of templated HB238 molecules is placed in a microcavity to achieve strong light-matter coupling between the cavity mode and the ordered molecular layer. The oriented transition dipole moments of the J- and H-transitions are schematically illustrated. The angle $\phi$ is between the orientation of the AGNRs and the plane of incidence. b) Polarization-resolved extinction spectra of a templated HB238 layer for normally incident light. The orientation of the polarizer with respect to the nominal direction of the graphene nanoribbons is indicated in the figure in 10\degree{} steps. c) Angle of incidence and polarization-resolved extinction spectra. The dark colored lines correspond to 45$^\circ$ AOI and the light colors to normal incidence. d) Atomic force micrograph of the HB238 surface. e) Polarization-resolved photoluminescence micrographs of HB238 films (left, polarization orthogonal to the AGNRs; right, polarization along the AGNRs).}\label{fig:concept}
\end{figure}

We explore light-matter interaction of ordered thin films of an organic semiconductor in a microcavity. The concept of the study is illustrated in Fig.~\ref{fig:concept}a. We use \ac{GNR}\cite{cai2010,senkovskiy2017} to template the growth of thin films of the merocyanine dye HB238. The optical and electrical properties of HB238 and similar merocyanines have been studied earlier.~\cite{Liess2017,liess2019,buerckstuemmer2011,Schembri2021,schaefer2024,boehner2024} HB238 forms molecular aggregates that show both a blue-shifted H-like and a red-shifted J-like optical transition with a photon energy difference of $\approx0.86$~eV relative to monomer. From hereon these transitions will be referred to as H- and J-transition, respectively. We previously reported on the optical properties of these aggregates in spin-coated thin films. In these films the \ac{TDM} of the J-transition is aligned along the substrate plane, while the \ac{TDM} of the H-transition is along the substrate normal (see Fig.~\ref{fig:concept}a).~\cite{schaefer2024,boehner2024} As a consequence, the J-transition can be excited with both s- and p-polarized light, while the H-transition can only be excited with p-polarized light and a non-zero \ac{AOI}. Due to oblique angle aggregation the \ac{TDM} of the J- and H-transitions are perpendicular to each other.~\cite{schaefer2024,boehner2024} 
The spectroscopic properties of the thin films can be understood within the theoretical framework developed by Kasha~\cite{kasha1963,kasha1965exciton} and Davydov~\cite{kasha1963,kasha1965exciton,davydov1964theory,davydov1971theory}. Molecular materials with more than two molecules per unit cell and oriented in an oblique fashion result in Davydov splitting with a red- (J-transition) and blue-shifted (H-transition) component. The \ac{TDM}s of the two transitions are orthogonal to each other.

Aligned \ac{GNR} can be used as a template to prepare films of HB238 via thermal \ac{PVD} to induce uniaxial growth of the molecular aggregates. For this, aligned 7-AGNRs are grown on a Au(788) crystal and transferred onto the bottom half of a microcavity using a previously published protocol.~\cite{senkovskiy2017,alavi2019} Full details of the sample fabrication are given in the Methods section. The transfer process maintains the alignment and quality of the nanoribbons. The alignment of the templating layer gets transferred to the molecules, so that HB238 molecules deposited on the surface are oriented along the ribbons. As a result, a material with properties resembling those of a biaxial crystal is formed. Here, the \ac{TDM} of the J-transition is aligned along the 7-AGNRs and that of the H-transition along the surface normal (see Fig.~\ref{fig:concept}a). We have recently shown that depositing HB238 thin films on crystalline substrates results in spectrally narrower optical transitions, significantly enlarged crystalline domains, and is expected to increase charge mobility.~\cite{boehner2024} These improvements in the material properties are attributed to reduced disorder. We expect that similar changes may be obtained also from templating using graphene nanoribbons.

The anisotropic structure of the templated thin film leads to an optical response that is dependent on the sample orientation and polarization of light. In Fig.~\ref{fig:concept}b the polarization-resolved extinction of a templated HB238 film is shown for normally incident light. Due to the normal incidence, only the red-shifted J-transition at approximately 1.65 eV photon energy is visible. The absorbance is maximized or minimized, if the polarization is parallel or perpendicular to the \ac{GNR}, respectively, demonstrating the successful alignment of the molecules. Using density functional theory we have shown earlier, that the J-transition is polarized along one crystal axis (axis b), while the H-transition is aligned along another crystal axis (axis a) orthogonal to the J-transition dipole moment.\cite{boehner2024}  Figure~\ref{fig:concept}c displays polarization-resolved extinction spectra for different angles of incidence (sample orientation $\phi~=~0$, see Fig.~\ref{fig:concept}a). We observe the proposed alignment of the \ac{TDM}s: The H-transition is only visible for p-polarized light and at non-normal incidence. The J-transition is visible independent of the angle of incidence. Due to the alignment of the sample (angle $\phi~=~0$), for p-polarized light the J-transition is observed only weakly due to the in-plane alignment of the molecules. The molecular alignment is further illustrated by the atomic force micrograph shown in Fig.~\ref{fig:concept}d. Clearly aligned linear structures are visible in the sample topography.

To gain further insight into the templating we perform laser-scanning confocal photoluminescence microscopy of the templated films. Figure~\ref{fig:concept}e shows polarization-resolved fluorescence micrographs (see details of the microscopy in the Methods section). We observe that for the polarization along the direction of the ribbons the micrograph shows a highly homogeneous emission with only minor variations. In contrast, for the orthogonal direction the variation in emission intensity is larger. The base intensity is lower than for the orthogonal polarization direction but there are brighter hotspots in the image. We attribute these to localized crystallites, which may either not be well oriented or in which the aggregation state is different from the overall structure of the films. However, fluorescence spectra collected at different locations show very similar spectra, suggesting that the aggregation state is homogeneous throughout the films (see the Supporting Information Fig.~S3). Crystallites can be observed in atomic force microscopy micrographs (see Fig.~\ref{fig:concept}d) suggesting that these are the origin of the hotspots. We conclude that the templated thin films have optical properties of a biaxial crystal with well oriented \ac{TDM}s of the J- and H-transitions.

\subsection*{Strongly coupled templated thin films}

\begin{figure}
    \centering
    \includegraphics[width=16cm]{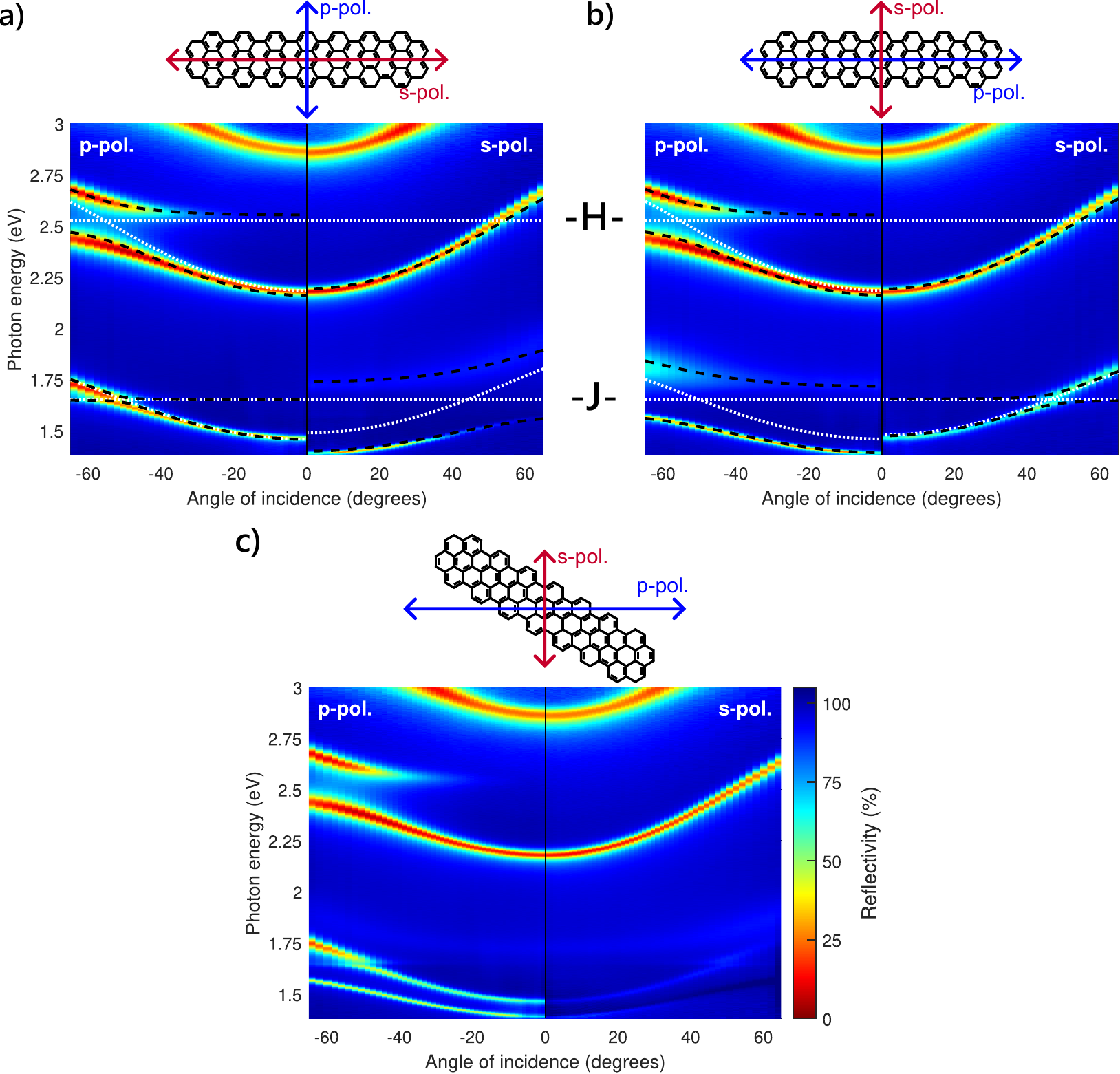}
    \caption{Angle-resolved reflectivity spectra of templated HB238 in a microcavity. a) Ribbons aligned parallel to electric field of s-polarized light ($\phi~=~0$). b)  Ribbons aligned parallel to electric field of p-polarized light ($\phi~=~\pi/2$). c) Angle-resolved reflectivity spectra with 45\degree{} ribbon orientation ($\phi~=~\pi/4$), showing polarization splitting of the polaritons. The fits to a coupled oscillator model are shown in black dashed lines. The spectral positions of the J- and H-transitions are shown with white dotted lines. The colorbar in panel c) is valid for all panels.}\label{fig:strong-coupling}
\end{figure}

We now turn to the optical properties of the biaxial templated thin film in a planar microcavity (see Fig~\ref{fig:strong-coupling}). We measure the angle-resolved reflection spectra of the samples for different sample orientation (angle $\phi$ in Fig.~\ref{fig:concept}a) and polarization of the incident light. For the case $\phi~=~0$ where the orientation of the 7-AGNRs is orthogonal to plane of incidence we observe strong-light matter interaction for the J- or H-transition depending on the polarization of the incident light as shown in Fig.~\ref{fig:strong-coupling}a. For p-polarized incident light we observe a clear anti-crossing around the H-transition at approximately 2.53~eV. This is in agreement with our earlier studies in which we showed that the \ac{TDM} of the H-transition is along the substrate normal (see Fig.~\ref{fig:concept}c).~\cite{schaefer2024} The in-plane anisotropy due to the 7-AGNRs does not appear to disturb this, and strong light-matter coupling is observed when the incident light has an electric field component along the surface normal. For the J-transition, there is a weak interaction visible in Fig.~\ref{fig:strong-coupling}a. However, the conditions for strong light-matter interaction are not fulfilled. The obtained Rabi-energy has a similar magnitude as the spectral width of the transition. For s-polarized incident light the interaction with the H-transition is absent and instead an avoided crossing around the J-transition is observed. In comparison to our previous report\cite{schaefer2024}, the fact that the coupling to the J-transition is absent for p-polarized incident light demonstrates that the \ac{TDM} of the J-transition is now oriented along the 7-AGNRs, a direction along which the electric field of the incident light has no component. This polarization-switching of the coupling is due to the biaxial nature of the film resulting from the templating with the ribbons. If the sample is rotated so that $\phi~=~\pi/2$ the coupling to both transitions is absent for s-polarized light while it is visible for both transitions for p-polarized light. This agrees excellently with the schematic of the orientation of the \ac{TDM}s shown in Fig.~\ref{fig:concept}a. Simulated reflection spectra reproduce the measurement well, see Fig.~S5 in the Supporting Information.

We fit the extracted positions of the minima in the reflection spectra using a coupled oscillator model (see details in the Methods section).~\cite{schaefer2024,Wu2022,skolnick1998strong} The fitted dispersion curves are shown with black dashed lines in Fig.~\ref{fig:strong-coupling}. The agreement of the fit to the data is excellent. From the fitted model we extract for the data in Fig.~\ref{fig:strong-coupling}a Rabi-energies of 299~meV for the J-polariton for s-polarized light and 190~meV for the H-polariton for p-polarized light. For p-polarized light the Rabi-energy for the J-polariton is only 26~meV and the system does not fulfill the conditions of strong-coupling. For the case shown in Fig.~\ref{fig:strong-coupling}b the Rabi-energies are slightly smaller and correspond to 263~meV and 189~meV for the J- and H-polaritons for p-polarized light, respectively. The Rabi-energy for the J-polariton for s-polarized light is 51~meV which is comparable to the spectral width of the transition and the system is not in the strong-coupling regime in this case. Comparing the Rabi-energies to the values for an almost identical system without templating studied earlier by us~\cite{schaefer2024}, we observe a significant increase as a consequence of the templating. 

We have so far considered only the high symmetry cases where the polarization of the incident light is along or orthogonal to the nanoribbons and thus to the aligned \ac{TDM} of the J-transition. The angle-resolved reflectivity spectrum for the case $\phi~=~\pi/4$ is shown in Fig~\ref{fig:strong-coupling}c. We now observe polarization mixing of the polaritons involving the J-transition and all polariton branches are visible for both p- and s-polarized light. The spectral features for the transitions involving the H-transition remain unchanged. This observation is similar to previous studies on and theoretical predictions for strong-light matter interaction involving single crystalline organic materials.~\cite{balagurov2004,kena-cohen2008,zoubi2005,litinskaya2004} 

\subsection*{Probing molecular order inside microcavity}

\begin{figure}
    \centering
    \includegraphics[width=16cm]{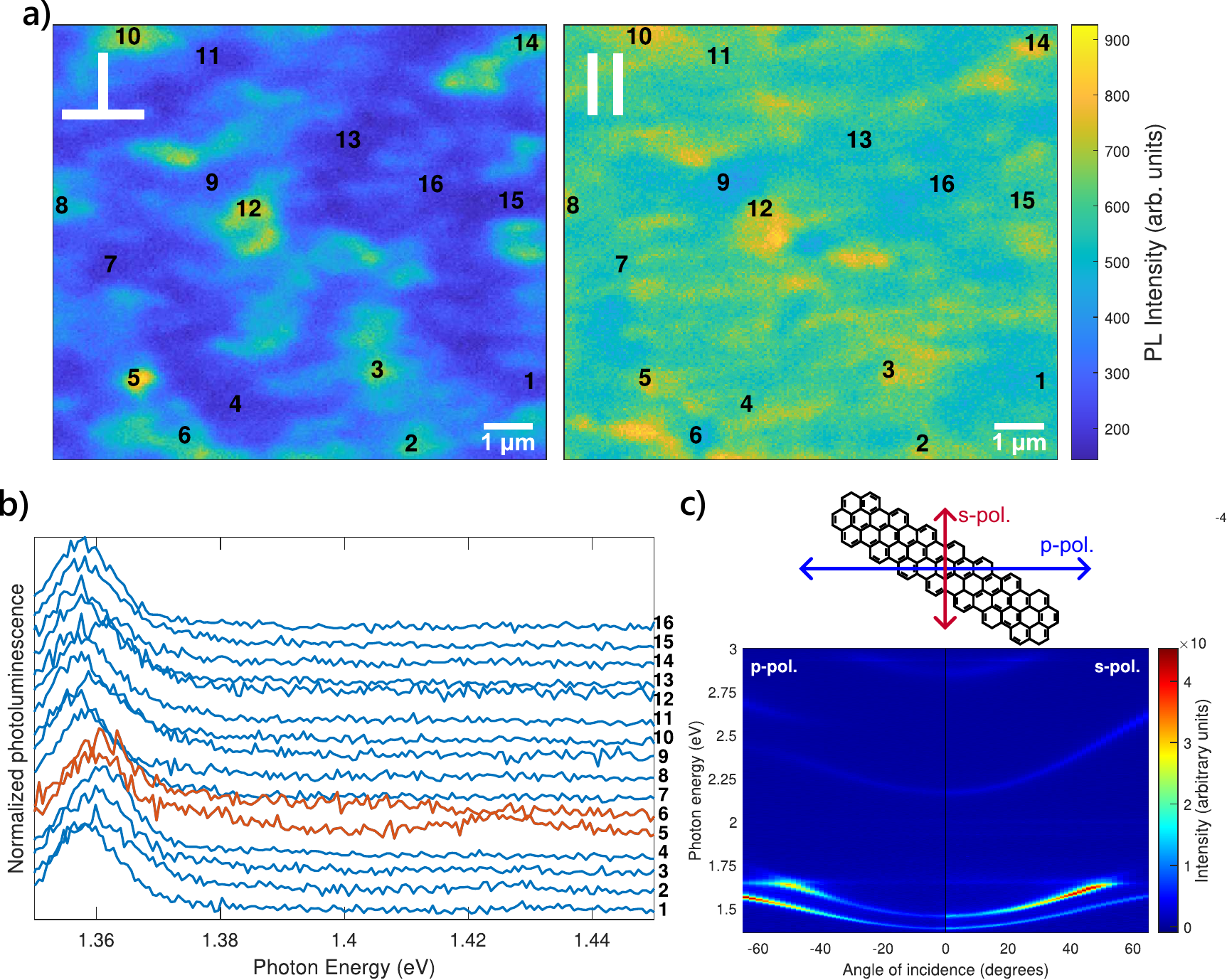}
    \caption{Polarization- and angle-resolved photoluminesence microscopy. a) Polarization-resolved photoluminescence micrographs of the aligned microcavity. The left (right) micrograph is for emission polarized orthogonal to (along) the graphene nanoribbons. b) Normalized emission spectra for light emitted along the sample normal. The numbers correspond to the locations shown in panel a at which the spectra were collected.  The polarization of the emitted light is filtered parallel to the ribbons. c) Angle-resolved emission spectra of templated HB238 in a microcavity. The ribbons are aligned 45\degree{} to the plane of incidence.}\label{fig:PL}
\end{figure}

The data visualized in Fig.~\ref{fig:strong-coupling} clearly demonstrates the orientational ordering of the excitonic component in our cavity and its influence on light-matter interaction. This allows selectively addressing different polariton bands using the polarization of light, angle of incidence, and orientation of sample. To probe the quality of the ordering, we next apply photoluminescence microspectroscopy (see details in the Methods section). If we orient the sample such that the angle $\phi$ is equal to $\pi/4$, all polariton branches involving the J-transition are excited by both p- and s-polarized light (see Fig.~\ref{fig:strong-coupling}c). This is also translated to the photoluminescence signal as shown in Fig.~\ref{fig:PL}c. Similar observations have been reported by \citet{kena-cohen2008} for an anthracene single crystal.  We observe that both lower polariton branches are bright independent of the polarization of the collected light. Surprisingly, different from reflection spectra (Fig.~\ref{fig:strong-coupling}), the p- and s-polarized emission are of almost equal brightness for the photoluminescence. 

The dependence of the photoluminescence spectrum on the polarization of light and the orientation of the sample (angle $\phi$) can be used to gain further insight into the quality of the orientational order of the excitonic component. For this purpose, we first perform laser scanning photoluminescence microscopy of the sample. Figure~\ref{fig:PL}a shows typical photoluminescence micrographs of the cavity for emitted light polarized orthogonal to (left panel) and along (right panel) the graphene nanoribbons. We observe that as expected the emission is much stronger for light polarized along the ribbons compared to the orthogonal case. However, the strength of the signal is not completely homogeneous but shows localized maxima. We assume that these originate from the crystallites in the templated HB238 thin film (see Fig.~\ref{fig:concept}d and e). For the emission from the cavity the inhomogeneities appear to be less significant than for the bare film (see Fig.~\ref{fig:concept}e). We speculate that this might be due to the coupling to the delocalized optical modes of the cavity. 

Next, we collect emission spectra at the locations numbered in Fig.~\ref{fig:PL}a. From the angle-resolved emission spectrum, Fig.~\ref{fig:PL}c, we conclude that for the orientation of the sample equal to $\phi~=~\pi/4$, the emission from the two lowest polaritons (J-transition) is approximately equal at normal incidence. As a matter of fact, only for the high-symmetry directions $\phi~=~0$ or $\phi~=~\pi/2$ we observe only one polariton branch, see Fig.~S4 in the Supporting Information. We can thus use the ratio of the emission from the two lower polariton branches of the J-transition as a measure for the quality of the alignment of the ribbons and its influence on the light-matter interaction. This has been theoretically investigated for crystalline materials.~\cite{litinskaya2004} We align the sample such that $\phi~=~\pi/2$ and measure s-polarized spectra, so that the electric field is in the plane of the TDM.  We record angle-resolved spectra at selected positions on the sample (see Fig.~\ref{fig:PL}a). The data for normally directed emission is shown in Fig.~\ref{fig:PL}b, extracted from back-focal plane measurements (see details in the Methods section). The spectra shown here have been normalized. The main conclusion from the data is that we observe almost identical emission spectra with only minor variations. The emission is dominated by the lowest polariton close to 1.36~eV photon energy. For two spectra a minor component of the other polariton is visible. We can thus conclude that the alignment of the excitonic component of our system is homogeneous and accurate and better than the intensity variations seen in Fig.~\ref{fig:concept}e and \ref{fig:PL}a. In contrast, in a region without the template the ratio between the two lowest polaritons varies strongly (see the Supporting Information, Fig.~S6). 

\subsection*{Influence of excitonic states on the polariton bands}

\begin{figure}
    \centering
    \includegraphics[width=15cm]{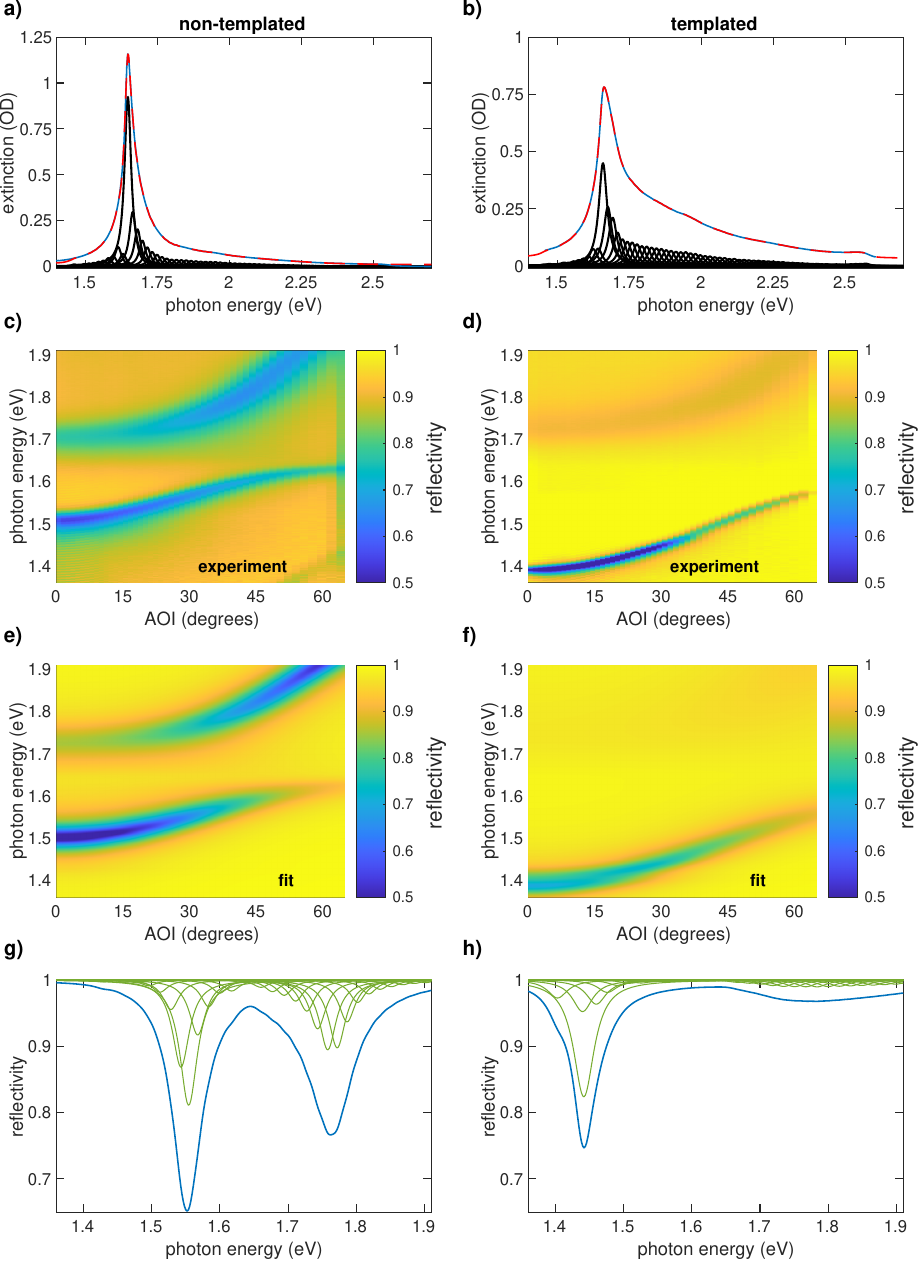}
    \caption{Multi-oscillator analysis of reflection spectra. a) and b) Fit to  extinction spectra. The blue solid line shows the measured extinction spectrum, and the red dashed line is the fitted model. The contributions of the individual Lorentzian oscillators are shown with black solid lines. The data for the non-templated and templated devices are shown in a and b, respectively. c) and d) Measured reflection spectra for the non-templated and templated devices, respectively. The corresponding fits are shown in e) and f). g) and h) The fitted reflection spectra at $30^\circ$ angle of incidence are shown with blue solid line. The contributions of the individual polaritons are shown with green solid lines. The data for the non-templated and templated devices are shown in g and h, respectively.}
    \label{fig:multi-oscillator}
\end{figure}

The presence of the graphene nanoribbons results in ordered growth of the excitonic component in our cavity. Our spectroscopic data clearly demonstrate anisotropic optical properties, which resemble those of single crystalline materials. We next turn to analyzing the polariton states of our system to see how they are influenced by the templating layer. We apply the formalism recently introduced by \citet{adom.202302387}. The key ingredient of this approach is accounting for inhomogeneous broadening of the (partially) disordered molecular system in the microcavity. We briefly review the formalism of Ref.~\cite{adom.202302387}. We represent the excitonic component as a set of Lorentzian oscillators. We fit $N$ oscillators with equally spaced peak energies $E_i = E_1+(i-1)D$ and identical spectral width $\Gamma$. Here $E_1$ is a constant photon energy representing the edge of of the considered spectral band, $D$ is the energy spacing between the oscillators, and the index $i=1,2,\ldots N$ numbers the oscillators. We fit the measured extinction at photon energy $E$ with the expression
\begin{equation}
    \sum_{i=1}^{N} \frac{A_i\left(\Gamma/2\right)^2}{\left(E-E_i\right)^2+\left(\Gamma/2\right)^2}.
\end{equation}
Here the amplitudes $A_i$ are used as fitting parameters. Figure~\ref{fig:multi-oscillator}a and b show the measured extinction spectra for non-templated (spincast) and templated HB238 thin films together with the fit result, respectively. The non-templated microcavity is almost identical to the templated device with potentially small differences due to fabrication errors. Full details for non-templated microcavities with HB238 can be found in Ref.~\cite{schaefer2024}. We observe that we can well describe the measured data with the multi-oscillator model and that the fit results are clearly different for the two cases. Here $N = 70$ oscillators were used. We remark that this choice is not critical and the conclusions are not changed, if this parameter is varied. The oblique-angle aggregation of HB238 results in bright J- and H-transitions, which are the bottom and top of the energy band of the material.\cite{Spano2007} The states between these are for an ideal aggregate dark. Due to disorder, however, the dark states couple to light. The spectra shown in Fig.~\ref{fig:multi-oscillator}a and b show that for the templated film many more oscillators couple to light than for the non-templated one. This suggests that, surprisingly, the templated material has more defects or uncoupled monomers than the amorphous one. This is unexpected considering the strongly polarization-dependent response of the templated thin film and our earlier observation of increase in crystal domain size on ordered surfaces~\cite{boehner2024}.

The Lorentzian oscillators are included in a matrix Hamiltonian commonly used to model strongly-coupled systems.\cite{adom.202302387,Wu2022,skolnick1998strong} The Hamiltonian $H$ takes the form
\begin{equation}
H = 
\begin{pmatrix}
E_{\mathrm{Photon}}(\theta)-i\Gamma_c & \hbar\Omega_1 & \hbar\Omega_2 & \ldots & \hbar \Omega_N \\
\hbar\Omega_1 & E_1-i\Gamma & 0 & \ldots & 0 \\
\hbar\Omega_2 & 0 & E_2-i\Gamma & \ldots & 0 \\
\vdots & \vdots & \vdots & \ddots & \vdots \\
\hbar \Omega_N & 0 & 0 & \ldots & E_N-i\Gamma
\end{pmatrix},
\label{hamiltonian}
\end{equation}
with 
\begin{equation}
E_{\mathrm{Photon}}(\theta) = \frac{hcm}{2 n_{\mathrm{cavity}} L}\left( 1 - \frac{\sin^{2}\theta}{n_{\mathrm{cavity}}^2}\right)^{-\frac{1}{2}},
\label{photon_energy}
\end{equation}
where $E_{\mathrm{Photon}}(\theta)$ is the cavity photon energy at angle of incidence $\theta$, and $h$ and $\hbar$ are the Planck and reduced Planck constant, respectively, $c$ is the speed of light in vacuum, $m$ is the cavity mode number, $n_{\mathrm{cavity}}$ is the effective refractive index of the cavity, and $L$ is the cavity length. In Eq.~\ref{hamiltonian}, the $E_i$ are the spectral positions of the Lorentzian oscillators. The light-matter coupling constants $\hbar \Omega_i$ are proportional to the square root of the amplitudes of the Lorentzians, $\sqrt{A_i}$. We remark that the constants $A_i$ describe the absorption, while the light-matter coupling depends on amplitude of the oscillators. This results in the square-root dependence. We have used one common proportionality constant between the $\hbar \Omega_i$ and $\sqrt{A_i}$ for all oscillators. We diagonalize the Hamiltonian Eq.~\ref{hamiltonian} to obtain the energy levels of the polaritons and the Hopfield coefficients. From the Hopfield coefficient describing the photon fraction of the polaritons, we obtain the reflection spectrum. The measured and fitted angle-resolved reflection spectra are shown in Fig.~\ref{fig:multi-oscillator}c-f. For the fitting we vary the proportionality constant between the coupling constant $\hbar \Omega_i$ and the amplitudes $\sqrt{A_i}$, the proportionality constant relating the photon fraction of a given polariton to the reflectivity, and the effective refractive index of the cavity $n_{\mathrm{cavity}}$. Using just these parameters we obtain a good agreement between the model and experimental data. The main differences are small discrepancies in the linewidths of the polaritons and the visibility of the upper polariton for the templated film.

The multi-oscillator model gives us access to the contribution of the individual oscillators to the observed polariton bands. In Fig.~\ref{fig:multi-oscillator}g and h we display the fitted reflection spectrum at $30^\circ$ angle of incidence (AOI) together with the contributions of the individual polaritons to the observed response. The situation is very different for the non-templated and templated devices. For the non-templated case, several polaritons contribute to the signal with comparable amplitudes. In contrast, for the templated cavity only a few polaritons contribute to the reflectivity dips. This is surprising in light of the spectrum of oscillators representing the excitonic component, Figs.~\ref{fig:multi-oscillator}a and b. Our interpretation of this result is that in the case of the templated film, one oscillator is dominantly responsible for the light-matter interaction. The other oscillators required to explain the extinction spectrum (Fig.~\ref{fig:multi-oscillator}b) originate from uncoupled monomers or defects. The templating thus appears to divide the material into two ensembles: an ordered component driven by the graphene nanoribbons and monomers or defects in the film. It thus appears that the order is increased by strongly coupling the aligned film to a microcavity mode. This is in agreement with a previous theoretical study by \citet{Spano2015}, where coupling a J-aggregate to a microcavity mode increased the coherence length.

\subsection*{Conclusions}

We have studied the light-matter interaction of a layer of HB238 merocyanine molecules templated with graphene nanoribbons. The templating results in a global alignment of the transition dipole moments of the J- and H-transitions of the aggregated thin film. The H-transition's transition dipole moment lies along the substrate normal while that of the J-transition is directed along the graphene nanoribbons. The anisotropic optical properties of the templated layer resemble those of a crystalline material. Our spectroscopic study demonstrates strong light-matter interaction of the templated merocyanine. The templating results in strongly polarization-dependent properties. We are able to address different polaritons using the polarization of light, the angle of incidence, and the orientation of the sample. The Rabi-energies extracted from the angle-resolved reflection spectra are significantly higher than those observed in earlier studies on similar systems with an spincast (non-templated) molecular layer. We attribute this to better alignment of the transition dipole moments with the electric field and possibly increased oscillator strength due to reduced disorder. 

Using a multi-oscillator model we conclude that templating significantly modifies the energy level structure of the organic semiconductor. For the templated microcavity, light-matter interaction is dominated by one polariton. Our data suggests that templating with graphene nanoribbons results in two ensembles of the excitonic component: One with increased order and a second much less ordered one containing defects or uncoupled monomers.

Using materials of reduced dimensionality could become a general and efficient way to order organic molecules. This has great potential for organic optoelectronics. The most immediate consequence of templating is anisotropic optical properties which allows using the polarization of light in engineering light-matter interaction. This is particularly interesting for optical metasurfaces~\cite{editorial2023,Yu2014} and in general optical nanostructures in which the intensity and polarization distributions can be engineered in detail. Combining such structures with templated organic layers has potential for novel and improved devices. The improved order in the templated layer also seems to result in modified energy level structure. Fully harvesting this is an exciting direction to pursue. In particular, the coherence and dynamics of the excitations in the material may be very different from the disordered counterpart.

\section*{Methods}

\subsection*{Sample fabrication}

\subsubsection*{Microcavities}
The microcavities were fabricated by thermal high vacuum physical vapor deposition (PVD) of 150~nm of silver (99.99\% purity) on glass substrates followed by 120~nm of silicon monoxide (99.99\% purity).  Next, a thin film of \textit{N},\textit{N}’-bis(4-(6-((3-ethyloxetan-3-yl)methoxy))-hexylphenyl)-\textit{N},\textit{N}’-diphenyl-4,4’-diamine (OTPD)~\cite{Zacharias2007} was spincast at 1000~rpm and an acceleration of 4000~rpm/s from a toluene solution containing 2~mg/ml OTPD and 0.5~wt.\% 4-octyloxydiphenyl-iodonium-hexafluoroantimonate (OPPI) under inert conditions in a glovebox, which yielded an approximately 17~nm thin film. The OTPD layer reduced the surface roughness of the sample. The OTPD film was then crosslinked by exposing the sample with ultra-violet light (365~nm) and subsequent annealing at 110~\degree{}C for one minute.~\cite{rudati_2012} The templating layer of graphene nanoribbons was transferred onto the OTPD layer and HB238 was subsequently deposited by thermal PVD. Details of this are given below. Afterwards 5~nm of silicon monoxide were evaporated, followed by a spincasting layer of polyvinyl alcohol (Mowiol 56-98, \textit{Sigma-Aldrich}) with an approximate thickness of 315~nm. The microcavity was closed by a 30~nm silver mirror deposited by PVD. 

\subsubsection*{7-AGNR synthesis and transfer and HB238 deposition}

Aligned 7-AGNRs were synthesised on Au(788) single crystals using a well-established synthesis route.~\cite{narita_2019,linden_2012} First 16~\AA~of 10,10’-dibromo-9,9’-bianthryl (DBBA) were deposited onto the clean Au(788) surface under UHV conditions. Heating to 200~$^\circ$C and 400~$^\circ$C resulted in 7-AGNRs. The GNR film was covered with a poly(methyl methacrylate) (PMMA) support layer and was subsequently delaminated from the gold surface using the bubble transfer technique.~\cite{sun_2016} After three washing steps, the GNR/PMMA film was placed on the target substrate (OTPD). After 24~h of drying, the PMMA layer was removed in a boiling acetone bath.

The HB238 was evaporated onto the GNR/OTPD substrate using a rate of 0.03~\AA/s and a substrate temperature of 72~$^\circ$C under high vacuum conditions (pressure approximately $10^{-7}$~mbar). The alignment of HB238 was investigated by atomic force microscopy using an MFP-3D Infinity AFM (\emph{Oxford Instruments Asylum Research}). The measurements were performed in alternating contact mode using AC200TS tips (\emph{Olympus}). The raw data were analysed using Gwyddion. 

\subsection*{Reference sample fabrication}
For the transmission spectra (Fig.~\ref{fig:concept}b,c) and photoluminescence micrographs (Fig.~\ref{fig:concept}e) a reference sample without microcavity was fabricated: OTPD was spincast on a quartz substrate and treated as described above. Graphene nanoribbons were transferred on the sample followed by evaporation of HB238. The fabrication of the templated layer for the reference sample were identical to that of the microcavities.

\subsection*{Absorbance  spectra}
A Lambda 1050 spectrometer (\textit{PerkinElmer}) with a three-detector module was used to measure absorbance spectra. The sample was rotated to record spectra at 45\degree{} AOI. 

\subsection*{Angle-resolved reflectivity and photoluminescence measurements}

The measurements were performed as in our earlier work Ref.~\cite{schaefer2024}. Briefly, light reflected or emitted from the sample is collected by a microscope lens (MPlanApo N, Olympus, numerical aperture 0.95). Two more lenses are positioned in a 4f-layout to image the back-focal plane of the objective on the spectrometer entrance slit (IsoPlaneSCT 320, \textit{Princeton Instruments}). The spectrometer is equipped with a deep-cooled charged coupled device  (Pixis, \textit{Princeton Instruments}). Light from a  halogen lamp was focused on the sample or on a silver mirror as reference for reflectivity measurements. Photoluminescence was measured with a focused laser beam with 405~nm wavelength (Chameleon Ultra II with Chameleon Compact OPO-Vis, \textit{Coherent}). The excitation was blocked from the collected light with a dielectric filter. To equally excite all linear polarized states, circularly polarized light was used for the excitation for all photoluminescence measurements. The reflected or emitted light was filtered with a broadband linear polarizer (UBB01A, \textit{Moxtek}) to analyze the polarization state.

\subsection*{Fluorescence microspectroscopy}

For the polarization-resolved photoluminescence micrographs in Fig.~\ref{fig:concept}e, the same setup and conditions as for the angle-resolved reflectivity and photoluminescence measurements was used. The emitted light passed through long-pass filter and the same polarizer as above, and was focused on a single-photon counting module (\textit{Micro Photon Devices Srl.}). The sample was scanned or positioned using a nanopositioning stage to acquire micrographs or spectra at selected locations. Photoluminescence spectra were recorded by the same spectrometer as above. 

For the polarization-resolved photoluminescence micrographs in Fig.~\ref{fig:PL}a the polarizer was removed and a second identical single-photon counting module was added. Two polarization components of the emitted light were collected simultaneously using the detectors and a polarizing beam splitter. The spectra in Fig.~\ref{fig:PL}b were recorded using the 4f configuration (details above) to obtain angle-resolved spectra. In the data analysis, the emission at 0\degree{} AOI was extracted.

\subsection*{Data analysis and simulations}

Reflectivity minima were identified with a peak finding function (Matlab, \textit{MathWorks}). We used a two coupled oscillator model to globally fit all dispersive curves for both s- and p-polarized light with the trust-region-reflective algorithm (Matlab, \textit{MathWorks}). Details on the fitting procedure and shared parameters fitting are given in Ref.~\cite{schaefer2024}. 

Finite element method (COMSOL Multiphysics with Wave Optics module in the frequency domain, \textit{COMSOL Multiphysics GmbH}) was employed to simulate the reflectivity spectra. We used a two-dimensional geometry with the complex refractive index for all materials (silver\cite{CIESIELSKI2017349}, silicon monoxide\cite{schaefer2024}, OTPD, HB238, polyvinyl alcohol). For OTPD, HB238, and polyvinyl alcohol material parameters determined using spectroscopic ellipsometry were used. Periodic Floquet boundary conditions were implemented along the substrate surface. 

\section*{Funding sources}

This project is funded with support from the RTG-2591 "TIDE - Template-designed Organic Electronics" (Deutsche Forschungsgemeinschaft). K. L. and R. S. acknowledge funding from the DFG project 426882575. Instrument funding by the Deutsche Forschungsgemeinschaft (project 448775637) in cooperation with the Ministerium für Kunst und Wissenschaft of North Rhine‐Westphalia is acknowledged.
	  
\section*{Acknowledgements}

Research was supported by the University of Cologne through the Institutional Strategy of the University of Cologne within the German Excellence Initiative (QM$^2$). We thank Stephanie R\"uth for synthesizing the OTPD and Dirk Hertel for helpful discussions.\\

\noindent The authors declare no competing financial interest.

\section*{Author contributions}

R.S. fabricated the microcavity samples, developed the measurement setup, carried out the optical measurements, and analyzed the data with support from K.L. P.W. prepared the templated HB238 films and performed atomic force microscopy. O.E. carried out ellipsometric characterization of templated HB238 films. K.M. participated in the supervision of the project. K.L. planned and supervised the project. All authors contributed to the analysis of the results and writing of the paper.
\begin{suppinfo}

Polarization-resolved extinction spectra for spincast and templated HB238 films, polar plot from exticntion spectra of templated HB238 films, photoluminescence spectra and micrographs for templated HB238 thin film, angle-resolved photoluminescence spectra of templated HB238 in a microcavity, simulated reflection spectra, photoluminescence spectra and micrographs of a microcavity with spincast and evaporated HB238.

\end{suppinfo}

\newpage
\input{literature.bbl}

\end{document}

%% file: literature.bbl
\providecommand{\latin}[1]{#1}
\makeatletter
\providecommand{\doi}
  {\begingroup\let\do\@makeother\dospecials
  \catcode`\{=1 \catcode`\}=2 \doi@aux}
\providecommand{\doi@aux}[1]{\endgroup\texttt{#1}}
\makeatother
\providecommand*\mcitethebibliography{\thebibliography}
\csname @ifundefined\endcsname{endmcitethebibliography}
  {\let\endmcitethebibliography\endthebibliography}{}

%% file: main_Arxiv_v2.bbl
\begin{mcitethebibliography}{70}
\providecommand*\natexlab[1]{#1}
\providecommand*\mciteSetBstSublistMode[1]{}
\providecommand*\mciteSetBstMaxWidthForm[2]{}
\providecommand*\mciteBstWouldAddEndPuncttrue
  {\def\EndOfBibitem{\unskip.}}
\providecommand*\mciteBstWouldAddEndPunctfalse
  {\let\EndOfBibitem\relax}
\providecommand*\mciteSetBstMidEndSepPunct[3]{}
\providecommand*\mciteSetBstSublistLabelBeginEnd[3]{}
\providecommand*\EndOfBibitem{}
\mciteSetBstSublistMode{f}
\mciteSetBstMaxWidthForm{subitem}{(\alph{mcitesubitemcount})}
\mciteSetBstSublistLabelBeginEnd
  {\mcitemaxwidthsubitemform\space}
  {\relax}
  {\relax}

\bibitem[Lidzey \latin{et~al.}(1998)Lidzey, Bradley, Skolnick, Virgili, Walker,
  and Whittaker]{Lidzey1998}
Lidzey,~D.~G.; Bradley,~D. D.~C.; Skolnick,~M.~S.; Virgili,~T.; Walker,~S.;
  Whittaker,~D.~M. {Strong exciton–photon coupling in an organic
  semiconductor microcavity}. \emph{Nature} \textbf{1998}, \emph{395}, 53\relax
\mciteBstWouldAddEndPuncttrue
\mciteSetBstMidEndSepPunct{\mcitedefaultmidpunct}
{\mcitedefaultendpunct}{\mcitedefaultseppunct}\relax
\EndOfBibitem
\bibitem[Zasedatelev \latin{et~al.}(2019)Zasedatelev, Baranikov, Urbonas,
  Scafirimuto, Scherf, Stöferle, Mahrt, and Lagoudakis]{Zasedatelev2019}
Zasedatelev,~A.~V.; Baranikov,~A.~V.; Urbonas,~D.; Scafirimuto,~F.; Scherf,~U.;
  Stöferle,~T.; Mahrt,~R.~F.; Lagoudakis,~P.~G. A room-temperature organic
  polariton transistor. \emph{Nature Photonics} \textbf{2019}, \emph{13},
  378--383\relax
\mciteBstWouldAddEndPuncttrue
\mciteSetBstMidEndSepPunct{\mcitedefaultmidpunct}
{\mcitedefaultendpunct}{\mcitedefaultseppunct}\relax
\EndOfBibitem
\bibitem[K\'ena-Cohen and Forrest(2010)K\'ena-Cohen, and
  Forrest]{Kena-Cohen2010}
K\'ena-Cohen,~S.; Forrest,~S.~R. Room-temperature polariton lasing in an
  organic single-crystal microcavity. \emph{Nature Photonics} \textbf{2010},
  \emph{4}, 371--375\relax
\mciteBstWouldAddEndPuncttrue
\mciteSetBstMidEndSepPunct{\mcitedefaultmidpunct}
{\mcitedefaultendpunct}{\mcitedefaultseppunct}\relax
\EndOfBibitem
\bibitem[Lerario \latin{et~al.}(2017)Lerario, Fieramosca, Barachati, Ballarini,
  Daskalakis, Dominici, Giorgi, Maier, Gigli, Kéna-Cohen, and
  Sanvitto]{Lerario2017}
Lerario,~G.; Fieramosca,~A.; Barachati,~F.; Ballarini,~D.; Daskalakis,~K.~S.;
  Dominici,~L.; Giorgi,~M.~D.; Maier,~S.~A.; Gigli,~G.; Kéna-Cohen,~S.;
  Sanvitto,~D. Room-temperature superfluidity in a polariton condensate.
  \emph{Nature Physics} \textbf{2017}, \emph{13}, 837--841\relax
\mciteBstWouldAddEndPuncttrue
\mciteSetBstMidEndSepPunct{\mcitedefaultmidpunct}
{\mcitedefaultendpunct}{\mcitedefaultseppunct}\relax
\EndOfBibitem
\bibitem[Daskalakis \latin{et~al.}(2014)Daskalakis, Maier, Murray, and
  Kéna-Cohen]{Daskalakis2014}
Daskalakis,~K.~S.; Maier,~S.~A.; Murray,~R.; Kéna-Cohen,~S. Nonlinear
  interactions in an organic polariton condensate. \emph{Nature Materials}
  \textbf{2014}, \emph{13}, 271--278\relax
\mciteBstWouldAddEndPuncttrue
\mciteSetBstMidEndSepPunct{\mcitedefaultmidpunct}
{\mcitedefaultendpunct}{\mcitedefaultseppunct}\relax
\EndOfBibitem
\bibitem[Plumhof \latin{et~al.}(2014)Plumhof, St{\"{o}}ferle, Mai, Scherf, and
  Mahrt]{Plumhof2014}
Plumhof,~J.~D.; St{\"{o}}ferle,~T.; Mai,~L.; Scherf,~U.; Mahrt,~R.~F.
  Room-temperature Bose--Einstein condensation of cavity exciton--polaritons in
  a polymer. \emph{Nature Materials} \textbf{2014}, \emph{13}, 247--252\relax
\mciteBstWouldAddEndPuncttrue
\mciteSetBstMidEndSepPunct{\mcitedefaultmidpunct}
{\mcitedefaultendpunct}{\mcitedefaultseppunct}\relax
\EndOfBibitem
\bibitem[Mischok \latin{et~al.}(2023)Mischok, Hillebrandt, Kwon, and
  Gather]{Mischok2023}
Mischok,~A.; Hillebrandt,~S.; Kwon,~S.; Gather,~M.~C. Highly efficient
  polaritonic light-emitting diodes with angle-independent narrowband emission.
  \emph{Nature Photonics} \textbf{2023}, \emph{17}, 393--400\relax
\mciteBstWouldAddEndPuncttrue
\mciteSetBstMidEndSepPunct{\mcitedefaultmidpunct}
{\mcitedefaultendpunct}{\mcitedefaultseppunct}\relax
\EndOfBibitem
\bibitem[Orgiu \latin{et~al.}(2015)Orgiu, George, Hutchison, Devaux, Dayen,
  Doudin, Stellacci, Genet, Schachenmayer, Genes, Pupillo, Samor{\`{i}}, and
  Ebbesen]{Orgiu2015}
Orgiu,~E.; George,~J.; Hutchison,~J.~A.; Devaux,~E.; Dayen,~J.~F.; Doudin,~B.;
  Stellacci,~F.; Genet,~C.; Schachenmayer,~J.; Genes,~C.; Pupillo,~G.;
  Samor{\`{i}},~P.; Ebbesen,~T.~W. {Conductivity in organic semiconductors
  hybridized with the vacuum field}. \emph{Nature Materials} \textbf{2015},
  \emph{14}, 1123--1129\relax
\mciteBstWouldAddEndPuncttrue
\mciteSetBstMidEndSepPunct{\mcitedefaultmidpunct}
{\mcitedefaultendpunct}{\mcitedefaultseppunct}\relax
\EndOfBibitem
\bibitem[Berghuis \latin{et~al.}(2022)Berghuis, Tichauer, de~Jong, Sokolovskii,
  Bai, Ramezani, Murai, Groenhof, and Gómez~Rivas]{berghuis2022}
Berghuis,~A.~M.; Tichauer,~R.~H.; de~Jong,~L. M.~A.; Sokolovskii,~I.; Bai,~P.;
  Ramezani,~M.; Murai,~S.; Groenhof,~G.; Gómez~Rivas,~J. Controlling Exciton
  Propagation in Organic Crystals through Strong Coupling to Plasmonic
  Nanoparticle Arrays. \emph{ACS Photonics} \textbf{2022}, \emph{9},
  2263--2272\relax
\mciteBstWouldAddEndPuncttrue
\mciteSetBstMidEndSepPunct{\mcitedefaultmidpunct}
{\mcitedefaultendpunct}{\mcitedefaultseppunct}\relax
\EndOfBibitem
\bibitem[Mischok \latin{et~al.}(2024)Mischok, Siegmund, Le~Roux, Hillebrandt,
  Vandewal, and Gather]{Mischok2024}
Mischok,~A.; Siegmund,~B.; Le~Roux,~F.; Hillebrandt,~S.; Vandewal,~K.;
  Gather,~M.~C. Breaking the angular dispersion limit in thin film optics by
  ultra-strong light-matter coupling. \emph{Nature Communications}
  \textbf{2024}, \emph{15}, 10529\relax
\mciteBstWouldAddEndPuncttrue
\mciteSetBstMidEndSepPunct{\mcitedefaultmidpunct}
{\mcitedefaultendpunct}{\mcitedefaultseppunct}\relax
\EndOfBibitem
\bibitem[W{\"{u}}rthner \latin{et~al.}(2011)W{\"{u}}rthner, Kaiser, and
  Saha-M{\"{o}}ller]{Wurthner2011}
W{\"{u}}rthner,~F.; Kaiser,~T.~E.; Saha-M{\"{o}}ller,~C.~R. {J-aggregates: from
  serendipitous discovery to supramolecular engineering of functional dye
  materials}. \emph{Angewandte Chemie International Edition} \textbf{2011},
  \emph{50}, 3376--3410\relax
\mciteBstWouldAddEndPuncttrue
\mciteSetBstMidEndSepPunct{\mcitedefaultmidpunct}
{\mcitedefaultendpunct}{\mcitedefaultseppunct}\relax
\EndOfBibitem
\bibitem[Fidder \latin{et~al.}(1990)Fidder, Knoester, and Wiersma]{Fidder1990}
Fidder,~H.; Knoester,~J.; Wiersma,~D.~A. {Superradiant emission and optical
  dephasing in J-aggregates}. \emph{Chemical Physics Letters} \textbf{1990},
  \emph{171}, 529--536\relax
\mciteBstWouldAddEndPuncttrue
\mciteSetBstMidEndSepPunct{\mcitedefaultmidpunct}
{\mcitedefaultendpunct}{\mcitedefaultseppunct}\relax
\EndOfBibitem
\bibitem[Kobayashi(1996)]{Kobayashi1996}
Kobayashi,~T. \emph{{J-Aggregates}}; World Scientific, 1996; p 240\relax
\mciteBstWouldAddEndPuncttrue
\mciteSetBstMidEndSepPunct{\mcitedefaultmidpunct}
{\mcitedefaultendpunct}{\mcitedefaultseppunct}\relax
\EndOfBibitem
\bibitem[Garcia-Vidal \latin{et~al.}(2021)Garcia-Vidal, Ciuti, and
  Ebbesen]{garcia-vidal:2021}
Garcia-Vidal,~F.~J.; Ciuti,~C.; Ebbesen,~T.~W. Manipulating matter by strong
  coupling to vacuum fields. \emph{Science} \textbf{2021}, \emph{373},
  eabd0336\relax
\mciteBstWouldAddEndPuncttrue
\mciteSetBstMidEndSepPunct{\mcitedefaultmidpunct}
{\mcitedefaultendpunct}{\mcitedefaultseppunct}\relax
\EndOfBibitem
\bibitem[Khazanov \latin{et~al.}(2023)Khazanov, Gunasekaran, George, Lomlu,
  Mukherjee, and Musser]{khazanov2023}
Khazanov,~T.; Gunasekaran,~S.; George,~A.; Lomlu,~R.; Mukherjee,~S.;
  Musser,~A.~J. Embrace the darkness: An experimental perspective on organic
  exciton–polaritons. \emph{Chemical Physics Reviews} \textbf{2023},
  \emph{4}, 041305\relax
\mciteBstWouldAddEndPuncttrue
\mciteSetBstMidEndSepPunct{\mcitedefaultmidpunct}
{\mcitedefaultendpunct}{\mcitedefaultseppunct}\relax
\EndOfBibitem
\bibitem[George \latin{et~al.}(2024)George, Geraghty, Kelsey, Mukherjee,
  Davidova, Kim, and Musser]{adom.202302387}
George,~A.; Geraghty,~T.; Kelsey,~Z.; Mukherjee,~S.; Davidova,~G.; Kim,~W.;
  Musser,~A.~J. Controlling the Manifold of Polariton States Through Molecular
  Disorder. \emph{Advanced Optical Materials} \textbf{2024}, \emph{12},
  2302387\relax
\mciteBstWouldAddEndPuncttrue
\mciteSetBstMidEndSepPunct{\mcitedefaultmidpunct}
{\mcitedefaultendpunct}{\mcitedefaultseppunct}\relax
\EndOfBibitem
\bibitem[Gera and Sebastian(2022)Gera, and Sebastian]{gera2022}
Gera,~T.; Sebastian,~K.~L. Effects of disorder on polaritonic and dark states
  in a cavity using the disordered Tavis–Cummings model. \emph{The Journal of
  Chemical Physics} \textbf{2022}, \emph{156}, 194304\relax
\mciteBstWouldAddEndPuncttrue
\mciteSetBstMidEndSepPunct{\mcitedefaultmidpunct}
{\mcitedefaultendpunct}{\mcitedefaultseppunct}\relax
\EndOfBibitem
\bibitem[Engelhardt and Cao(2022)Engelhardt, and Cao]{engelhardt2022}
Engelhardt,~G.; Cao,~J. Unusual dynamical properties of disordered polaritons
  in microcavities. \emph{Phys. Rev. B} \textbf{2022}, \emph{105}, 064205\relax
\mciteBstWouldAddEndPuncttrue
\mciteSetBstMidEndSepPunct{\mcitedefaultmidpunct}
{\mcitedefaultendpunct}{\mcitedefaultseppunct}\relax
\EndOfBibitem
\bibitem[Agranovich \latin{et~al.}(2003)Agranovich, Litinskaia, and
  Lidzey]{agranovich2003}
Agranovich,~V.~M.; Litinskaia,~M.; Lidzey,~D.~G. Cavity polaritons in
  microcavities containing disordered organic semiconductors. \emph{Phys. Rev.
  B} \textbf{2003}, \emph{67}, 085311\relax
\mciteBstWouldAddEndPuncttrue
\mciteSetBstMidEndSepPunct{\mcitedefaultmidpunct}
{\mcitedefaultendpunct}{\mcitedefaultseppunct}\relax
\EndOfBibitem
\bibitem[Litinskaya and Reineker(2006)Litinskaya, and Reineker]{litinskaya2006}
Litinskaya,~M.; Reineker,~P. Loss of coherence of exciton polaritons in
  inhomogeneous organic microcavities. \emph{Phys. Rev. B} \textbf{2006},
  \emph{74}, 165320\relax
\mciteBstWouldAddEndPuncttrue
\mciteSetBstMidEndSepPunct{\mcitedefaultmidpunct}
{\mcitedefaultendpunct}{\mcitedefaultseppunct}\relax
\EndOfBibitem
\bibitem[Spano(2015)]{Spano2015}
Spano,~F.~C. Optical microcavities enhance the exciton coherence length and
  eliminate vibronic coupling in J-aggregates. \emph{The Journal of Chemical
  Physics} \textbf{2015}, \emph{142}, 184707\relax
\mciteBstWouldAddEndPuncttrue
\mciteSetBstMidEndSepPunct{\mcitedefaultmidpunct}
{\mcitedefaultendpunct}{\mcitedefaultseppunct}\relax
\EndOfBibitem
\bibitem[K\'ena-Cohen \latin{et~al.}(2008)K\'ena-Cohen,
  Davan\ifmmode~\mbox{\c{c}}\else \c{c}\fi{}o, and Forrest]{kena-cohen2008}
K\'ena-Cohen,~S.; Davan\ifmmode~\mbox{\c{c}}\else \c{c}\fi{}o,~M.;
  Forrest,~S.~R. Strong Exciton-Photon Coupling in an Organic Single Crystal
  Microcavity. \emph{Phys. Rev. Lett.} \textbf{2008}, \emph{101}, 116401\relax
\mciteBstWouldAddEndPuncttrue
\mciteSetBstMidEndSepPunct{\mcitedefaultmidpunct}
{\mcitedefaultendpunct}{\mcitedefaultseppunct}\relax
\EndOfBibitem
\bibitem[Berghuis \latin{et~al.}(2022)Berghuis, Tichauer, de~Jong, Sokolovskii,
  Bai, Ramezani, Murai, Groenhof, and Gómez~Rivas]{berghuis2022b}
Berghuis,~A.~M.; Tichauer,~R.~H.; de~Jong,~L. M.~A.; Sokolovskii,~I.; Bai,~P.;
  Ramezani,~M.; Murai,~S.; Groenhof,~G.; Gómez~Rivas,~J. Controlling Exciton
  Propagation in Organic Crystals through Strong Coupling to Plasmonic
  Nanoparticle Arrays. \emph{ACS Photonics} \textbf{2022}, \emph{9},
  2263--2272\relax
\mciteBstWouldAddEndPuncttrue
\mciteSetBstMidEndSepPunct{\mcitedefaultmidpunct}
{\mcitedefaultendpunct}{\mcitedefaultseppunct}\relax
\EndOfBibitem
\bibitem[Gärtner \latin{et~al.}(2014)Gärtner, Fiedler, Bauer, Marele, and
  Sokolowski]{Gaertner2014}
Gärtner,~S.; Fiedler,~B.; Bauer,~O.; Marele,~A.; Sokolowski,~M.~M. Lateral
  ordering of PTCDA on the clean and the oxygen pre-covered Cu(100) surface
  investigated by scanning tunneling microscopy and low energy electron
  diffraction. \emph{Beilstein Journal of Organic Chemistry} \textbf{2014},
  \emph{10}, 2055--2064\relax
\mciteBstWouldAddEndPuncttrue
\mciteSetBstMidEndSepPunct{\mcitedefaultmidpunct}
{\mcitedefaultendpunct}{\mcitedefaultseppunct}\relax
\EndOfBibitem
\bibitem[Trembułowicz \latin{et~al.}(2021)Trembułowicz, Sabik, and
  Grodzicki]{molecules26082393}
Trembułowicz,~A.; Sabik,~A.; Grodzicki,~M. Au(100) as a Template for Pentacene
  Monolayer. \emph{Molecules} \textbf{2021}, \emph{26}\relax
\mciteBstWouldAddEndPuncttrue
\mciteSetBstMidEndSepPunct{\mcitedefaultmidpunct}
{\mcitedefaultendpunct}{\mcitedefaultseppunct}\relax
\EndOfBibitem
\bibitem[Guo \latin{et~al.}(2014)Guo, Wang, Du, and Gao]{Guo_2014}
Guo,~H.; Wang,~Y.; Du,~S.; Gao,~H.-j. High-resolution scanning tunneling
microscopy imaging of Si(111) 7$\times$7 structure and intrinsic
  molecular states. \emph{Journal of Physics: Condensed Matter} \textbf{2014},
  \emph{26}, 394001\relax
\mciteBstWouldAddEndPuncttrue
\mciteSetBstMidEndSepPunct{\mcitedefaultmidpunct}
{\mcitedefaultendpunct}{\mcitedefaultseppunct}\relax
\EndOfBibitem
\bibitem[Öcal \latin{et~al.}(2025)Öcal, Weitkamp, Meerholz, and
  Olthof]{OCAL2025122690}
Öcal,~B.; Weitkamp,~P.; Meerholz,~K.; Olthof,~S. Chemical interaction and
  molecular growth of a highly dipolar merocyanine molecule on metal surfaces:
  A photoelectron spectroscopy study. \emph{Surface Science} \textbf{2025},
  \emph{754}, 122690\relax
\mciteBstWouldAddEndPuncttrue
\mciteSetBstMidEndSepPunct{\mcitedefaultmidpunct}
{\mcitedefaultendpunct}{\mcitedefaultseppunct}\relax
\EndOfBibitem
\bibitem[Bürckstümmer \latin{et~al.}(2011)Bürckstümmer, Tulyakova,
  Deppisch, Lenze, Kronenberg, Gsänger, Stolte, Meerholz, and
  Würthner]{buerckstuemmer2011}
Bürckstümmer,~H.; Tulyakova,~E.~V.; Deppisch,~M.; Lenze,~M.~R.;
  Kronenberg,~N.~M.; Gsänger,~M.; Stolte,~M.; Meerholz,~K.; Würthner,~F.
  Efficient solution-processed bulk heterojunction solar cells by antiparallel
  supramolecular arrangement of dipolar donor--acceptor dyes. \emph{Angewandte
  Chemie International Edition} \textbf{2011}, \emph{50}, 11628--11632\relax
\mciteBstWouldAddEndPuncttrue
\mciteSetBstMidEndSepPunct{\mcitedefaultmidpunct}
{\mcitedefaultendpunct}{\mcitedefaultseppunct}\relax
\EndOfBibitem
\bibitem[Kny \latin{et~al.}(2023)Kny, Reimer, Al-Shamery, Tomar, Bredow,
  Olthof, Hertel, Meerholz, and Sokolowski]{D3NR00767G}
Kny,~A.~J.; Reimer,~M.; Al-Shamery,~N.; Tomar,~R.; Bredow,~T.; Olthof,~S.;
  Hertel,~D.; Meerholz,~K.; Sokolowski,~M. Chiral self-organized single
  2D-layers of tetramers from a functional donor–acceptor molecule by the
  surface template effect. \emph{Nanoscale} \textbf{2023}, \emph{15},
  10319--10329\relax
\mciteBstWouldAddEndPuncttrue
\mciteSetBstMidEndSepPunct{\mcitedefaultmidpunct}
{\mcitedefaultendpunct}{\mcitedefaultseppunct}\relax
\EndOfBibitem
\bibitem[Kratzer \latin{et~al.}(2019)Kratzer, Matkovic, and
  Teichert]{Kratzer_2019}
Kratzer,~M.; Matkovic,~A.; Teichert,~C. Adsorption and epitaxial growth of
  small organic semiconductors on hexagonal boron nitride. \emph{Journal of
  Physics D: Applied Physics} \textbf{2019}, \emph{52}, 383001\relax
\mciteBstWouldAddEndPuncttrue
\mciteSetBstMidEndSepPunct{\mcitedefaultmidpunct}
{\mcitedefaultendpunct}{\mcitedefaultseppunct}\relax
\EndOfBibitem
\bibitem[Hlawacek and Teichert(2013)Hlawacek, and Teichert]{Hlawacek_2013}
Hlawacek,~G.; Teichert,~C. Nucleation and growth of thin films of rod-like
  conjugated molecules. \emph{Journal of Physics: Condensed Matter}
  \textbf{2013}, \emph{25}, 143202\relax
\mciteBstWouldAddEndPuncttrue
\mciteSetBstMidEndSepPunct{\mcitedefaultmidpunct}
{\mcitedefaultendpunct}{\mcitedefaultseppunct}\relax
\EndOfBibitem
\bibitem[Kratzer and Teichert(2016)Kratzer, and Teichert]{Kratzer_2016}
Kratzer,~M.; Teichert,~C. Thin film growth of aromatic rod-like molecules on
  graphene. \emph{Nanotechnology} \textbf{2016}, \emph{27}, 292001\relax
\mciteBstWouldAddEndPuncttrue
\mciteSetBstMidEndSepPunct{\mcitedefaultmidpunct}
{\mcitedefaultendpunct}{\mcitedefaultseppunct}\relax
\EndOfBibitem
\bibitem[Giri \latin{et~al.}(2014)Giri, Li, Smilgies, Li, Diao, Lenn, Chiu,
  Lin, Allen, Reinspach, Mannsfeld, Thoroddsen, Clancy, Bao, and
  Amassian]{Giri_2014}
Giri,~G.; Li,~R.; Smilgies,~D.-M.; Li,~E.~Q.; Diao,~Y.; Lenn,~K.~M.; Chiu,~M.;
  Lin,~D.~W.; Allen,~R.; Reinspach,~J.; Mannsfeld,~S. C.~B.; Thoroddsen,~S.~T.;
  Clancy,~P.; Bao,~Z.; Amassian,~A. One-dimensional self-confinement promotes
  polymorph selection in large-area organic semiconductor thin films.
  \emph{Nature Communications} \textbf{2014}, \emph{5}, 3573\relax
\mciteBstWouldAddEndPuncttrue
\mciteSetBstMidEndSepPunct{\mcitedefaultmidpunct}
{\mcitedefaultendpunct}{\mcitedefaultseppunct}\relax
\EndOfBibitem
\bibitem[Schembri \latin{et~al.}(2024)Schembri, Kolb, Stolte, and
  Würthner]{D4TC00678J}
Schembri,~T.; Kolb,~L.; Stolte,~M.; Würthner,~F. Polarized{,} color-selective
  and semi-transparent organic photodiode of aligned merocyanine H-aggregates.
  \emph{J. Mater. Chem. C} \textbf{2024}, \emph{12}, 4948--4953\relax
\mciteBstWouldAddEndPuncttrue
\mciteSetBstMidEndSepPunct{\mcitedefaultmidpunct}
{\mcitedefaultendpunct}{\mcitedefaultseppunct}\relax
\EndOfBibitem
\bibitem[Shaw \latin{et~al.}(2016)Shaw, Hayoz, Diao, Reinspach, To, Toney,
  Weitz, and Bao]{shaw_2016}
Shaw,~L.; Hayoz,~P.; Diao,~Y.; Reinspach,~J.~A.; To,~J. W.~F.; Toney,~M.~F.;
  Weitz,~R.~T.; Bao,~Z. Direct Uniaxial Alignment of a Donor--Acceptor
  Semiconducting Polymer Using Single-Step Solution Shearing. \emph{ACS Applied
  Materials \& Interfaces} \textbf{2016}, \emph{8}, 9285--9296\relax
\mciteBstWouldAddEndPuncttrue
\mciteSetBstMidEndSepPunct{\mcitedefaultmidpunct}
{\mcitedefaultendpunct}{\mcitedefaultseppunct}\relax
\EndOfBibitem
\bibitem[Herrmann \latin{et~al.}(2024)Herrmann, Hertzog, Mischok, Gather, and
  Zaumseil]{Herrmann2024}
Herrmann,~N.~J.; Hertzog,~M.; Mischok,~A.; Gather,~M.~C.; Zaumseil,~J.
  Polarization-Dependent Strong and Weak Light-Matter Coupling in Aligned
  Perylene Diimide Thin Films. \emph{ACS Applied Optical Materials}
  \textbf{2024}, \emph{2}, 1619--1628\relax
\mciteBstWouldAddEndPuncttrue
\mciteSetBstMidEndSepPunct{\mcitedefaultmidpunct}
{\mcitedefaultendpunct}{\mcitedefaultseppunct}\relax
\EndOfBibitem
\bibitem[Le~Roux \latin{et~al.}(2020)Le~Roux, Taylor, and Bradley]{LeRoux2020}
Le~Roux,~F.; Taylor,~R.~A.; Bradley,~D. D.~C. Enhanced and
  Polarization-Dependent Coupling for Photoaligned Liquid Crystalline
  Conjugated Polymer Microcavities. \emph{ACS Photonics} \textbf{2020},
  \emph{7}, 746--758\relax
\mciteBstWouldAddEndPuncttrue
\mciteSetBstMidEndSepPunct{\mcitedefaultmidpunct}
{\mcitedefaultendpunct}{\mcitedefaultseppunct}\relax
\EndOfBibitem
\bibitem[Gr{\"o}ning \latin{et~al.}(2018)Gr{\"o}ning, Wang, Yao, Pignedoli,
  Borin~Barin, Daniels, Cupo, Meunier, Feng, Narita, M{\"u}llen, Ruffieux, and
  Fasel]{groening2018}
Gr{\"o}ning,~O.; Wang,~S.; Yao,~X.; Pignedoli,~C.~A.; Borin~Barin,~G.;
  Daniels,~C.; Cupo,~A.; Meunier,~V.; Feng,~X.; Narita,~A.; M{\"u}llen,~K.;
  Ruffieux,~P.; Fasel,~R. Engineering of robust topological quantum phases in
  graphene nanoribbons. \emph{Nature} \textbf{2018}, \emph{560}, 209--213\relax
\mciteBstWouldAddEndPuncttrue
\mciteSetBstMidEndSepPunct{\mcitedefaultmidpunct}
{\mcitedefaultendpunct}{\mcitedefaultseppunct}\relax
\EndOfBibitem
\bibitem[Rizzo \latin{et~al.}(2018)Rizzo, Veber, Cao, Bronner, Chen, Zhao,
  Rodriguez, Louie, Crommie, and Fischer]{rizzo2018}
Rizzo,~D.~J.; Veber,~G.; Cao,~T.; Bronner,~C.; Chen,~T.; Zhao,~F.;
  Rodriguez,~H.; Louie,~S.~G.; Crommie,~M.~F.; Fischer,~F.~R. Topological band
  engineering of graphene nanoribbons. \emph{Nature} \textbf{2018}, \emph{560},
  204--208\relax
\mciteBstWouldAddEndPuncttrue
\mciteSetBstMidEndSepPunct{\mcitedefaultmidpunct}
{\mcitedefaultendpunct}{\mcitedefaultseppunct}\relax
\EndOfBibitem
\bibitem[Ruffieux \latin{et~al.}(2016)Ruffieux, Wang, Yang,
  S{\'a}nchez-S{\'a}nchez, Liu, Dienel, Talirz, Shinde, Pignedoli, Passerone,
  Dumslaff, Feng, M{\"u}llen, and Fasel]{ruffieux2016}
Ruffieux,~P.; Wang,~S.; Yang,~B.; S{\'a}nchez-S{\'a}nchez,~C.; Liu,~J.;
  Dienel,~T.; Talirz,~L.; Shinde,~P.; Pignedoli,~C.~A.; Passerone,~D.;
  Dumslaff,~T.; Feng,~X.; M{\"u}llen,~K.; Fasel,~R. On-surface synthesis of
  graphene nanoribbons with zigzag edge topology. \emph{Nature} \textbf{2016},
  \emph{531}, 489--492\relax
\mciteBstWouldAddEndPuncttrue
\mciteSetBstMidEndSepPunct{\mcitedefaultmidpunct}
{\mcitedefaultendpunct}{\mcitedefaultseppunct}\relax
\EndOfBibitem
\bibitem[Llinas \latin{et~al.}(2017)Llinas, Fairbrother, Borin~Barin, Shi, Lee,
  Wu, Yong~Choi, Braganza, Lear, Kau, Choi, Chen, Pedramrazi, Dumslaff, Narita,
  Feng, M{\"u}llen, Fischer, Zettl, Ruffieux, Yablonovitch, Crommie, Fasel, and
  Bokor]{llinas2017}
Llinas,~J.~P. \latin{et~al.}  Short-channel field-effect transistors with
  9-atom and 13-atom wide graphene nanoribbons. \emph{Nature Communications}
  \textbf{2017}, \emph{8}, 633\relax
\mciteBstWouldAddEndPuncttrue
\mciteSetBstMidEndSepPunct{\mcitedefaultmidpunct}
{\mcitedefaultendpunct}{\mcitedefaultseppunct}\relax
\EndOfBibitem
\bibitem[Ruffieux \latin{et~al.}(2012)Ruffieux, Cai, Plumb, Patthey, Prezzi,
  Ferretti, Molinari, Feng, M{\"u}llen, Pignedoli, and Fasel]{ruffieux2012}
Ruffieux,~P.; Cai,~J.; Plumb,~N.~C.; Patthey,~L.; Prezzi,~D.; Ferretti,~A.;
  Molinari,~E.; Feng,~X.; M{\"u}llen,~K.; Pignedoli,~C.~A.; Fasel,~R.
  Electronic Structure of Atomically Precise Graphene Nanoribbons. \emph{ACS
  Nano} \textbf{2012}, \emph{6}, 6930--6935\relax
\mciteBstWouldAddEndPuncttrue
\mciteSetBstMidEndSepPunct{\mcitedefaultmidpunct}
{\mcitedefaultendpunct}{\mcitedefaultseppunct}\relax
\EndOfBibitem
\bibitem[Denk \latin{et~al.}(2014)Denk, Hohage, Zeppenfeld, Cai, Pignedoli,
  S{\"o}de, Fasel, Feng, M{\"u}llen, Wang, Prezzi, Ferretti, Ruini, Molinari,
  and Ruffieux]{denk2014}
Denk,~R.; Hohage,~M.; Zeppenfeld,~P.; Cai,~J.; Pignedoli,~C.~A.; S{\"o}de,~H.;
  Fasel,~R.; Feng,~X.; M{\"u}llen,~K.; Wang,~S.; Prezzi,~D.; Ferretti,~A.;
  Ruini,~A.; Molinari,~E.; Ruffieux,~P. Exciton-dominated optical response of
  ultra-narrow graphene nanoribbons. \emph{Nature Communications}
  \textbf{2014}, \emph{5}, 4253\relax
\mciteBstWouldAddEndPuncttrue
\mciteSetBstMidEndSepPunct{\mcitedefaultmidpunct}
{\mcitedefaultendpunct}{\mcitedefaultseppunct}\relax
\EndOfBibitem
\bibitem[Cai \latin{et~al.}(2010)Cai, Ruffieux, Jaafar, Bieri, Braun,
  Blankenburg, Muoth, Seitsonen, Saleh, Feng, M{\"u}llen, and Fasel]{cai2010}
Cai,~J.; Ruffieux,~P.; Jaafar,~R.; Bieri,~M.; Braun,~T.; Blankenburg,~S.;
  Muoth,~M.; Seitsonen,~A.~P.; Saleh,~M.; Feng,~X.; M{\"u}llen,~K.; Fasel,~R.
  Atomically precise bottom-up fabrication of graphene nanoribbons.
  \emph{Nature} \textbf{2010}, \emph{466}, 470--473\relax
\mciteBstWouldAddEndPuncttrue
\mciteSetBstMidEndSepPunct{\mcitedefaultmidpunct}
{\mcitedefaultendpunct}{\mcitedefaultseppunct}\relax
\EndOfBibitem
\bibitem[Senkovskiy \latin{et~al.}(2017)Senkovskiy, Pfeiffer, Alavi, Bliesener,
  Zhu, Michel, Fedorov, German, Hertel, Haberer, Petaccia, Fischer, Meerholz,
  van Loosdrecht, Lindfors, and Gr{\"u}neis]{senkovskiy2017}
Senkovskiy,~B.~V. \latin{et~al.}  Making Graphene Nanoribbons Photoluminescent.
  \emph{Nano Letters} \textbf{2017}, \emph{17}, 4029--4037\relax
\mciteBstWouldAddEndPuncttrue
\mciteSetBstMidEndSepPunct{\mcitedefaultmidpunct}
{\mcitedefaultendpunct}{\mcitedefaultseppunct}\relax
\EndOfBibitem
\bibitem[Liess \latin{et~al.}(2017)Liess, Lv, Arjona-Esteban, Bialas, Krause,
  Stepanenko, Stolte, and Würthner]{Liess2017}
Liess,~A.; Lv,~A.; Arjona-Esteban,~A.; Bialas,~D.; Krause,~A.-M.;
  Stepanenko,~V.; Stolte,~M.; Würthner,~F. Exciton Coupling of Merocyanine
  Dyes from H- to J-type in the Solid State by Crystal Engineering. \emph{Nano
  Letters} \textbf{2017}, \emph{17}, 1719--1726\relax
\mciteBstWouldAddEndPuncttrue
\mciteSetBstMidEndSepPunct{\mcitedefaultmidpunct}
{\mcitedefaultendpunct}{\mcitedefaultseppunct}\relax
\EndOfBibitem
\bibitem[Liess \latin{et~al.}(2019)Liess, Arjona-Esteban, Kudzus, Albert,
  Krause, Lv, Stolte, Meerholz, and Würthner]{liess2019}
Liess,~A.; Arjona-Esteban,~A.; Kudzus,~A.; Albert,~J.; Krause,~A.-M.; Lv,~A.;
  Stolte,~M.; Meerholz,~K.; Würthner,~F. Ultranarrow bandwidth organic
  photodiodes by exchange narrowing in merocyanine H-and J-aggregate excitonic
  systems. \emph{Advanced Functional Materials} \textbf{2019}, \emph{29},
  1805058\relax
\mciteBstWouldAddEndPuncttrue
\mciteSetBstMidEndSepPunct{\mcitedefaultmidpunct}
{\mcitedefaultendpunct}{\mcitedefaultseppunct}\relax
\EndOfBibitem
\bibitem[Schembri \latin{et~al.}(2021)Schembri, Kim, Liess, Stepanenko, Stolte,
  and Würthner]{Schembri2021}
Schembri,~T.; Kim,~J.~H.; Liess,~A.; Stepanenko,~V.; Stolte,~M.; Würthner,~F.
  Semitransparent Layers of Social Self-Sorting Merocyanine Dyes for
  Ultranarrow Bandwidth Organic Photodiodes. \emph{Advanced Optical Materials}
  \textbf{2021}, \emph{9}, 2100213\relax
\mciteBstWouldAddEndPuncttrue
\mciteSetBstMidEndSepPunct{\mcitedefaultmidpunct}
{\mcitedefaultendpunct}{\mcitedefaultseppunct}\relax
\EndOfBibitem
\bibitem[Sch\"afer \latin{et~al.}(2024)Sch\"afer, B\"ohner, Schiek, Hertel,
  Meerholz, and Lindfors]{schaefer2024}
Sch\"afer,~R.; B\"ohner,~L.; Schiek,~M.; Hertel,~D.; Meerholz,~K.; Lindfors,~K.
  Strong Light--Matter Interaction of Molecular Aggregates with Two Excitonic
  Transitions. \emph{ACS Photonics} \textbf{2024}, \emph{11}, 111--120\relax
\mciteBstWouldAddEndPuncttrue
\mciteSetBstMidEndSepPunct{\mcitedefaultmidpunct}
{\mcitedefaultendpunct}{\mcitedefaultseppunct}\relax
\EndOfBibitem
\bibitem[Böhner \latin{et~al.}(2025)Böhner, Weitkamp, Limböck, Gildemeister,
  Fazzi, Schiek, Bruker, Hertel, Schäfer, Lindfors, and Meerholz]{boehner2024}
Böhner,~L.; Weitkamp,~P.; Limböck,~T.; Gildemeister,~N.; Fazzi,~D.;
  Schiek,~M.; Bruker,~R.; Hertel,~D.; Schäfer,~R.; Lindfors,~K.; Meerholz,~K.
  Influencing optical and charge transport properties by controlling the
  molecular interactions of merocyanine thin films. \emph{Org. Chem. Front.}
  \textbf{2025}, --\relax
\mciteBstWouldAddEndPuncttrue
\mciteSetBstMidEndSepPunct{\mcitedefaultmidpunct}
{\mcitedefaultendpunct}{\mcitedefaultseppunct}\relax
\EndOfBibitem
\bibitem[Kasha(1963)]{kasha1963}
Kasha,~M. Energy transfer mechanisms and the molecular exciton model for
  molecular aggregates. \emph{Radiation Research} \textbf{1963}, \emph{20},
  55--70\relax
\mciteBstWouldAddEndPuncttrue
\mciteSetBstMidEndSepPunct{\mcitedefaultmidpunct}
{\mcitedefaultendpunct}{\mcitedefaultseppunct}\relax
\EndOfBibitem
\bibitem[Kasha \latin{et~al.}(1965)Kasha, Rawls, and
  El-Bayoumi]{kasha1965exciton}
Kasha,~M.; Rawls,~H.~R.; El-Bayoumi,~M.~A. The exciton model in molecular
  spectroscopy. \emph{Pure and applied Chemistry} \textbf{1965}, \emph{11},
  371--392\relax
\mciteBstWouldAddEndPuncttrue
\mciteSetBstMidEndSepPunct{\mcitedefaultmidpunct}
{\mcitedefaultendpunct}{\mcitedefaultseppunct}\relax
\EndOfBibitem
\bibitem[Davydov(1964)]{davydov1964theory}
Davydov,~A.~S. The theory of molecular excitons. \emph{Soviet Physics Uspekhi}
  \textbf{1964}, \emph{7}, 145\relax
\mciteBstWouldAddEndPuncttrue
\mciteSetBstMidEndSepPunct{\mcitedefaultmidpunct}
{\mcitedefaultendpunct}{\mcitedefaultseppunct}\relax
\EndOfBibitem
\bibitem[Davydov and Dresner(1971)Davydov, and Dresner]{davydov1971theory}
Davydov,~A.~S.; Dresner,~S. \emph{Theory of molecular excitons}; Plenum Press,
  New York, 1971\relax
\mciteBstWouldAddEndPuncttrue
\mciteSetBstMidEndSepPunct{\mcitedefaultmidpunct}
{\mcitedefaultendpunct}{\mcitedefaultseppunct}\relax
\EndOfBibitem
\bibitem[Alavi \latin{et~al.}(2019)Alavi, Senkovskiy, Pfeiffer, Haberer,
  Fischer, Grueneis, and Lindfors]{alavi2019}
Alavi,~S.~K.; Senkovskiy,~B.,~V; Pfeiffer,~M.; Haberer,~D.; Fischer,~F.~R.;
  Grueneis,~A.; Lindfors,~K. Probing the origin of photoluminescence
  brightening in graphene nanoribbons. \emph{2D Materials} \textbf{2019},
  \emph{6}\relax
\mciteBstWouldAddEndPuncttrue
\mciteSetBstMidEndSepPunct{\mcitedefaultmidpunct}
{\mcitedefaultendpunct}{\mcitedefaultseppunct}\relax
\EndOfBibitem
\bibitem[Wu \latin{et~al.}(2022)Wu, Finkelstein-Shapiro, Wang, Rosenkampff,
  Yartsev, Pascher, Nguyen-Phan, Cogdell, Börjesson, and Pullerits]{Wu2022}
Wu,~F.; Finkelstein-Shapiro,~D.; Wang,~M.; Rosenkampff,~I.; Yartsev,~A.;
  Pascher,~T.; Nguyen-Phan,~T.~C.; Cogdell,~R.; Börjesson,~K.; Pullerits,~T.
  Optical cavity-mediated exciton dynamics in photosynthetic light harvesting 2
  complexes. \emph{Nature Communications} \textbf{2022}, \emph{13}, 6864\relax
\mciteBstWouldAddEndPuncttrue
\mciteSetBstMidEndSepPunct{\mcitedefaultmidpunct}
{\mcitedefaultendpunct}{\mcitedefaultseppunct}\relax
\EndOfBibitem
\bibitem[Skolnick \latin{et~al.}(1998)Skolnick, Fisher, and
  Whittaker]{skolnick1998strong}
Skolnick,~M.~S.; Fisher,~T.~A.; Whittaker,~D.~M. Strong coupling phenomena in
  quantum microcavity structures. \emph{Semiconductor Science and Technology}
  \textbf{1998}, \emph{13}, 645\relax
\mciteBstWouldAddEndPuncttrue
\mciteSetBstMidEndSepPunct{\mcitedefaultmidpunct}
{\mcitedefaultendpunct}{\mcitedefaultseppunct}\relax
\EndOfBibitem
\bibitem[Balagurov and La~Rocca(2004)Balagurov, and La~Rocca]{balagurov2004}
Balagurov,~D.~B.; La~Rocca,~G.~C. Organic microcavities with anisotropic
  optically active materials. \emph{physica status solidi (c)} \textbf{2004},
  \emph{1}, 518--521\relax
\mciteBstWouldAddEndPuncttrue
\mciteSetBstMidEndSepPunct{\mcitedefaultmidpunct}
{\mcitedefaultendpunct}{\mcitedefaultseppunct}\relax
\EndOfBibitem
\bibitem[Zoubi and La~Rocca(2005)Zoubi, and La~Rocca]{zoubi2005}
Zoubi,~H.; La~Rocca,~G.~C. Microscopic theory of anisotropic organic cavity
  exciton polaritons. \emph{Phys. Rev. B} \textbf{2005}, \emph{71},
  235316\relax
\mciteBstWouldAddEndPuncttrue
\mciteSetBstMidEndSepPunct{\mcitedefaultmidpunct}
{\mcitedefaultendpunct}{\mcitedefaultseppunct}\relax
\EndOfBibitem
\bibitem[Litinskaya \latin{et~al.}(2004)Litinskaya, Reineker, and
  Agranovich]{litinskaya2004}
Litinskaya,~M.; Reineker,~P.; Agranovich,~V.~M. Exciton–polaritons in a
  crystalline anisotropic organic microcavity. \emph{physica status solidi (a)}
  \textbf{2004}, \emph{201}, 646--654\relax
\mciteBstWouldAddEndPuncttrue
\mciteSetBstMidEndSepPunct{\mcitedefaultmidpunct}
{\mcitedefaultendpunct}{\mcitedefaultseppunct}\relax
\EndOfBibitem
\bibitem[Spano \latin{et~al.}(2007)Spano, Silvestri, Spearman, Raimondo, and
  Tavazzi]{Spano2007}
Spano,~F.~C.; Silvestri,~L.; Spearman,~P.; Raimondo,~L.; Tavazzi,~S.
  {Reclassifying exciton-phonon coupling in molecular aggregates: Evidence of
  strong nonadiabatic coupling in oligothiophene crystals}. \emph{Journal of
  Chemical Physics} \textbf{2007}, \emph{127}, 184703\relax
\mciteBstWouldAddEndPuncttrue
\mciteSetBstMidEndSepPunct{\mcitedefaultmidpunct}
{\mcitedefaultendpunct}{\mcitedefaultseppunct}\relax
\EndOfBibitem
\bibitem[edi(2023)]{editorial2023}
Metasurfaces go mainstream. \emph{Nature Photonics} \textbf{2023}, \emph{17},
  1--1\relax
\mciteBstWouldAddEndPuncttrue
\mciteSetBstMidEndSepPunct{\mcitedefaultmidpunct}
{\mcitedefaultendpunct}{\mcitedefaultseppunct}\relax
\EndOfBibitem
\bibitem[Yu and Capasso(2014)Yu, and Capasso]{Yu2014}
Yu,~N.; Capasso,~F. Flat optics with designer metasurfaces. \emph{Nature
  Materials} \textbf{2014}, \emph{13}, 139--150\relax
\mciteBstWouldAddEndPuncttrue
\mciteSetBstMidEndSepPunct{\mcitedefaultmidpunct}
{\mcitedefaultendpunct}{\mcitedefaultseppunct}\relax
\EndOfBibitem
\bibitem[Zacharias \latin{et~al.}(2007)Zacharias, Gather, Rojahn, Nuyken, and
  Meerholz]{Zacharias2007}
Zacharias,~P.; Gather,~M.~C.; Rojahn,~M.; Nuyken,~O.; Meerholz,~K. New
  Crosslinkable Hole Conductors for Blue‐Phosphorescent Organic
  Light‐Emitting Diodes. \emph{Angewandte Chemie International Edition}
  \textbf{2007}, \emph{46}, 4388--4392\relax
\mciteBstWouldAddEndPuncttrue
\mciteSetBstMidEndSepPunct{\mcitedefaultmidpunct}
{\mcitedefaultendpunct}{\mcitedefaultseppunct}\relax
\EndOfBibitem
\bibitem[Rudati \latin{et~al.}(2012)Rudati, Mueller, and Meerholz]{rudati_2012}
Rudati,~P.; Mueller,~D.; Meerholz,~K. Preparation of Insoluble Hole-Injection
  Layers by Cationic Ring-Opening Polymerisation of Oxetane-Derivatized
  TriPhenylamineDimer for Organic Electronics Devices. \emph{Procedia
  Chemistry} \textbf{2012}, \emph{4}, 216--223, The International Conference on
  Innovation in Polymer Science and Technology\relax
\mciteBstWouldAddEndPuncttrue
\mciteSetBstMidEndSepPunct{\mcitedefaultmidpunct}
{\mcitedefaultendpunct}{\mcitedefaultseppunct}\relax
\EndOfBibitem
\bibitem[Narita \latin{et~al.}(2019)Narita, Chen, Chen, and
  Müllen]{narita_2019}
Narita,~A.; Chen,~Z.; Chen,~Q.; Müllen,~K. Solution and on-surface synthesis
  of structurally defined graphene nanoribbons as a new family of
  semiconductors. \emph{Chem. Sci.} \textbf{2019}, \emph{10}, 964--975\relax
\mciteBstWouldAddEndPuncttrue
\mciteSetBstMidEndSepPunct{\mcitedefaultmidpunct}
{\mcitedefaultendpunct}{\mcitedefaultseppunct}\relax
\EndOfBibitem
\bibitem[Linden \latin{et~al.}(2012)Linden, Zhong, Timmer, Aghdassi, Franke,
  Zhang, Feng, M\"ullen, Fuchs, Chi, and Zacharias]{linden_2012}
Linden,~S.; Zhong,~D.; Timmer,~A.; Aghdassi,~N.; Franke,~J.~H.; Zhang,~H.;
  Feng,~X.; M\"ullen,~K.; Fuchs,~H.; Chi,~L.; Zacharias,~H. Electronic
  Structure of Spatially Aligned Graphene Nanoribbons on Au(788). \emph{Phys.
  Rev. Lett.} \textbf{2012}, \emph{108}, 216801\relax
\mciteBstWouldAddEndPuncttrue
\mciteSetBstMidEndSepPunct{\mcitedefaultmidpunct}
{\mcitedefaultendpunct}{\mcitedefaultseppunct}\relax
\EndOfBibitem
\bibitem[Sun \latin{et~al.}(2016)Sun, Deng, Guo, Zhan, Deng, Xu, Fan, Xu, Guo,
  Huang, and Liu]{sun_2016}
Sun,~J.; Deng,~S.; Guo,~W.; Zhan,~Z.; Deng,~J.; Xu,~C.; Fan,~X.; Xu,~K.;
  Guo,~W.; Huang,~Y.; Liu,~X. Electrochemical Bubbling Transfer of Graphene
  Using a Polymer Support with Encapsulated Air Gap as Permeation Stopping
  Layer. \emph{Journal of Nanomaterials} \textbf{2016}, \emph{2016},
  7024246\relax
\mciteBstWouldAddEndPuncttrue
\mciteSetBstMidEndSepPunct{\mcitedefaultmidpunct}
{\mcitedefaultendpunct}{\mcitedefaultseppunct}\relax
\EndOfBibitem
\bibitem[Ciesielski \latin{et~al.}(2017)Ciesielski, Skowronski, Trzcinski, and
  Szoplik]{CIESIELSKI2017349}
Ciesielski,~A.; Skowronski,~L.; Trzcinski,~M.; Szoplik,~T. Controlling the
  optical parameters of self-assembled silver films with wetting layers and
  annealing. \emph{Applied Surface Science} \textbf{2017}, \emph{421},
  349--356\relax
\mciteBstWouldAddEndPuncttrue
\mciteSetBstMidEndSepPunct{\mcitedefaultmidpunct}
{\mcitedefaultendpunct}{\mcitedefaultseppunct}\relax
\EndOfBibitem
\end{mcitethebibliography}
